\documentclass[11pt,a4paper,reqno]{amsart}
\usepackage[T1]{fontenc}
\usepackage{lmodern}
\usepackage{amsmath,amssymb,amsthm,mathtools}
\usepackage[a4paper,margin=28mm]{geometry}
\usepackage{microtype}
\usepackage{enumitem}
\usepackage{hyperref}
\usepackage{xcolor}
\hypersetup{hidelinks,pdftitle={Complete entropy contraction from conditional hyperboundedness}}
\DeclareMathOperator{\Tr}{Tr}
\DeclareMathOperator{\id}{id}
\DeclareMathOperator{\supp}{supp}

\newtheorem{theorem}{Theorem}[section]
\newtheorem{proposition}[theorem]{Proposition}
\newtheorem{lemma}[theorem]{Lemma}
\newtheorem{corollary}[theorem]{Corollary}
\theoremstyle{definition}

\theoremstyle{remark}
\newtheorem{remark}[theorem]{Remark}

\numberwithin{equation}{section}
\allowdisplaybreaks[1]
\setlist[enumerate]{leftmargin=*,itemsep=3pt}
\title{Complete entropy contraction from conditional hyperboundedness}
\author{Li Gao and Lijun Wang}
\date{}

\begin{document}
\begin{abstract}
For finite-dimensional quantum channels compatible with a faithful
conditional expectation, we show that conditional hyperboundedness and strict
GNS contraction together yield a relative entropy contraction. For KMS-symmetric
quantum Markov semigroups, a conditional
$2\to q$ hyperboundedness at time $t_0>0$ combined with a GNS gap $\lambda>0$ implies a modified
logarithmic Sobolev constant 
\[ \alpha=\frac{\lambda(1-2/q)}
 {12(\lambda t_0+\log M+1-2/q)},\]
where $M$ is the hyperboundedness constant. As a consequence, completely bounded hyperboundedness over auxiliary
systems gives complete versions of entropy contraction with the same rates.
\end{abstract}
\maketitle

\section{Introduction}
\label{sec:introduction}

Relative entropy is a fundamental information measure of distinguishability between two quantum states \cite{Umegaki62,HiaiPetz91}. A key property of quantum relative entropy is the
data-processing inequality, which states that any quantum information processing modeled by quantum channels cannot increase
this distinguishability \cite{Lindblad75,Frenkel}. It is then natural to quantify the loss of information via a strong data-processing inequality that asks for a contraction coefficient strictly below one \cite{HRF22}. For quantum Markov semigroups, the
corresponding exponential entropy decay is characterized by a modified
logarithmic Sobolev inequality (MLSI), which is a powerful tool for obtaining quantitative bounds on
mixing time to equilibrium \cite{KT13,Bardet17}. When considering the correlated initial states with an external reference system, the complete
versions of entropy contraction were studied for the same contraction rate after adjoining any idle
auxiliary system, uniformly in its dimension \cite{Frenkel,GR,GJL}.

In the classical framework of Markov chains, detailed balance is a standard condition to ensure the equilibrium state. In the quantum setting, the detailed balance condition admits different generalizations.
Under Gelfand--Naimark--Segal (GNS) detailed balance, the strongest among the quantum detailed balance conditions, a general route
from spectral estimates to complete entropy decay is available. Gao and
Rouz\'e \cite{GR} established finite-dimensional complete MLSI bounds in
terms of the spectral gap and the complete index of the equilibrium
expectation. Gao, Junge, LaRacuente, and Li \cite{GJLL} later obtained
logarithmic index dependence through mixing in complete positivity
order. More recently, Gao and Guo \cite{GG,GGBKM} obtained channel
contraction and complete entropy-decay criteria without detailed
balance, using GNS contraction or coercivity. These results motivate a natural question: whether the index assumption, which can be seen as an ultracontractive estimate, can be replaced by a weaker regularity estimate in terms of hyperboundedness.

A further motivation comes from the recent breakthrough of Chen, Kastoryano,
and Gily\'en \cite{CKG23}, who constructed an efficiently implementable
Gibbs sampler satisfying exact Kubo--Martin--Schwinger (KMS)
detailed balance for noncommuting Hamiltonians. Their construction,
and its extensions \cite{DLL25}, make KMS-symmetric Lindbladians
central to quantum thermal state preparation. KMS symmetry is weaker
than GNS symmetry, so the GNS-symmetric entropy bounds do not apply
directly. Subsequent work proved high-temperature rapid mixing
\cite{RSFA26}, size-uniform KMS spectral gaps for one-dimensional spin
chains at every fixed finite temperature \cite{BC25}, and further
mixing and entropy-decay estimates for one-dimensional noncommuting
systems \cite{RSF26,CPS,CD,GGW26} as well as at high temperature in general spatial dimensions \cite{CDL26}.
Liu, Wan, and Wu \cite{LWWEntropic} proved MLSI for every
finite-dimensional primitive KMS-symmetric quantum Markov semigroup
and gave a primitive KMS-symmetric example without CMLSI.
These
developments call for entropy estimates that can use the available
norm and gap information beyond GNS detailed balance.

In the primitive KMS-symmetric setting, hyperboundedness offers such
a connection. A single finite-time $L_2$-to-$L_q$ estimate, together
with a spectral gap, gives quantitative hypercontractive and
$L_2$-logarithmic Sobolev bounds \cite{TPK14}.
Wan and Wu \cite{WWCheeger} established a quantum Cheeger inequality
for finite-dimensional primitive KMS-symmetric semigroups and obtained
quantitative logarithmic Sobolev and hypercontractivity bounds.
In our previous work
\cite{GW}, we obtained a direct passage from hyperboundedness and
centered $L_2$ contraction to entropy contraction for a single channel,
and derived MLSI bounds for primitive KMS-symmetric semigroups without
assuming $L_p$-regularity. Becker, Rouz\'e, and Salzmann \cite{BRS}
extended the positive-time criterion to an infinite-dimensional
primitive KMS setting.
For finite-dimensional primitive KMS-symmetric semigroups, Wu
\cite{WuReverse} proved reverse hypercontractivity for
$1/2<p\le q<1$ and showed that the $p=1/2$ logarithmic Sobolev
constant can vanish.

The non-primitive case presents an additional obstruction. The limiting
state is $E_*\rho$, where $E$ is the equilibrium conditional
expectation, and therefore depends on the input. Bardet
\cite{Bardet17} developed the corresponding MLSI framework, and
Bardet and Rouz\'e \cite{BR} formulated hypercontractivity and weak
logarithmic Sobolev inequalities using amalgamated $L_p$ norms. In
this setting the defect-free $L_2$-logarithmic Sobolev inequality often
fails \cite[Theorem~5.1]{BR}, even though entropy decay can hold.
This is related to the fact that the Rothaus
lemma used to remove a logarithmic Sobolev defect in the primitive case has no direct
non-primitive extension \cite{GJL}: the obstruction also occurs for matrix
amplifications of symmetric classical semigroups
\cite[Proposition~1.5]{GJLL}. Thus the primitive route through
defect removal to defect-free hypercontractivity or $L_2$-LSI does
not extend to the conditional and complete setting for quantum systems.

Inspired by our primitive results, we study a direct passage from
conditional hyperboundedness to entropy contraction with the help
of a gap condition. The defect is removed at the entropy level.
The resulting rates depend only on the
hyperboundedness data and gap, which
replaces the order control used in \cite{GG}.

We present our main results below. 
Throughout, $\mathcal H$ is a finite-dimensional Hilbert space, $\Tr$ is the ordinary
trace, and logarithms are natural. Write
$D(\rho\|\sigma)=\Tr\rho(\log\rho-\log\sigma)$ as the relative entropy with the usual support
convention. We use the maps $\Phi$ and $E$ in the Heisenberg picture and write
$\Phi_*$ and $E_*$ for their trace preduals.
For a faithful state $\sigma$, the weighted $L_p$-norm and GNS $L_2$-norm satisfy
$$\|X\|_{p,\sigma}
=\|\sigma^{1/(2p)}X\sigma^{1/(2p)}\|_{p}, \qquad 
\|X\|_{2,\mathrm{GNS},\sigma}^2=\Tr\sigma X^*X.$$
At $p=2$, the weighted norm is the KMS norm.
The conditional $L_p$-spaces $L_2^q(\mathcal N,\sigma)$ are defined in
Section~\ref{sec:prelim}. We note that the KMS norm used in hyperboundedness
and the GNS norm used in the coercivity condition play distinct
roles.

\begin{theorem}
\label{thm:main}Let $\Phi:B(\mathcal H)\to B(\mathcal H)$ be a unital completely positive map with a faithful invariant state $\sigma=\Phi_*\sigma$, and let
$E:B(\mathcal H)\to\mathcal N$ be the $\sigma$-preserving conditional
expectation onto the fixed-point subalgebra $\mathcal N$. Suppose that
\begin{enumerate}
\item[i)]  \emph{(conditional hyperboundedness)} for some
$2<q<\infty$ and $M\ge1$\begin{equation}
 \|\Phi:L_2(\sigma)
 \longrightarrow L_2^q(\mathcal N,\sigma)\|\le M.
 \label{eq:hb}
\end{equation}
\item[ii)] \emph{(GNS coercivity)} for some $0\le s<1$,
\begin{equation}
 \|(\Phi-E)X\|_{2,\mathrm{GNS},\sigma}
 \le s\|X\|_{2,\mathrm{GNS},\sigma}
 \qquad(X\in B(\mathcal H)).
 \label{eq:gns}
\end{equation}
\end{enumerate}
Then there is an explicit
$\varepsilon=\varepsilon(q,M,s)>0$ such that, for every
state $\rho$,
\begin{equation}
 D(\Phi_*\rho\|E_*\rho)
 \le(1-\varepsilon)D(\rho\|E_*\rho).
 \label{eq:main-ec}
\end{equation}
In particular, for each $q>2$, we have
\begin{equation}
 \varepsilon(q,M,s)\ge c(q)\frac{1-s^2}{1+\log M},
 \label{eq:fixed-q-summary}
\end{equation}
with the explicit constant
$c(q)\sim\frac{\pi}{128}\frac{q-2}{\log(1/(q-2))}$ as $q\downarrow2$.
\end{theorem}

An explicit choice of $c(q)>0$ is given in~\eqref{eq:cq-explicit}, which has no
dependence on dimension, the index, or the number of blocks in the fixed algebra. In the primitive case $\mathcal N=\mathbb CI$, GNS
contraction implies the centered KMS bound, and a similar entropy contraction estimate was already obtained in 
\cite[Theorem~2.4]{GW}. Our theorem treats non-trivial expectations using conditional $L_p$-norms and yields entropy contraction with respect to the input-dependent reference state.

A fixed-time application of Theorem~\ref{thm:main} to a semigroup
does not by itself give the exponential decay of relative entropy, so-called modified log-Sobolev inequality. To obtain entropy decay right from time zero, we use a
small-time conditional hyperboundedness curve, which, under KMS symmetry, can be obtained via
a single positive-time hyperboundedness estimate
by complex interpolation. Combined with GNS coercivity and the
entropy--energy comparison, it gives the following continuous-time result.
KMS symmetry here means self-adjointness for the KMS inner product
$\langle X,Y\rangle_{\mathrm{KMS},\sigma}
=\Tr X^*\sigma^{1/2}Y\sigma^{1/2}$.

\begin{theorem}\label{thm:kms-cmlsi}
Let $T_t=e^{t\mathcal L}$ be a quantum Markov semigroup that is
KMS-symmetric with respect to a faithful invariant state $\sigma$, and let $E$ be the $\sigma$-preserving
conditional expectation onto the fixed subalgebra $\mathcal N$. 
Suppose that 
\begin{enumerate}
\item \emph{(conditional hyperboundedness)} for some $t_0>0$, $2<q\le \infty$, and $M\ge1$,
\begin{equation}\label{eq:kms-single-time-hb}
 \|T_{t_0}:L_2(\sigma)
 \longrightarrow L_2^q(\mathcal N,\sigma)\|\le M.
\end{equation}
\item \emph{(GNS gap)} there exists
$\lambda>0$ such that
\begin{equation}\label{eq:generator-gns-gap}
\lambda :=\inf_{\substack{EX=0\\X\ne0}} \frac{ -\Re\Tr\sigma X^*\mathcal L(X)}{\|X\|_{2,\mathrm{GNS},\sigma}^2}>0
\end{equation}
\end{enumerate}
Then, for every $t\ge0$ and state $\rho$,
\begin{equation}\label{eq:kms-cmlsi-decay}
 D(T_{t*}\rho\|E_*\rho)
 \le e^{-2\alpha t}D(\rho\|E_*\rho),
\end{equation}
where $\alpha$ can be chosen as\begin{equation}\label{eq:kms-cmlsi-rate}
 \alpha=\frac1{12}\frac{\lambda (1-2/q)}{\lambda t_0+\log M+(1-2/q)}.
\end{equation}
\end{theorem}

For primitive semigroups, \cite[Theorem~3.3]{GW} gives the same
positive-time parameter scale, up to absolute constants, using
ordinary hyperboundedness and the KMS spectral gap (see \cite{BRS} for infinite-dimensional cases). Here conditional
norms allow the non-primitive case with a general fixed-point algebra. A separate
zero-time embedding bound of $\mathrm{id}:L_2\to L_2^\infty$ recovers logarithmic-index estimates
without symmetry in \cite{GG}; see Corollary~\ref{cor:zero-time-embedding} and
Remark~\ref{rem:zero-time-index}.

Note that the GNS gap and coercivity bound are preserved under semigroup amplification $T_t\otimes \id_{R}$ over any reference system $\mathcal H_R$
\cite[Proposition~5.3]{GGBKM}. Hence, applying Theorem~\ref{thm:kms-cmlsi} under a completely bounded
hyperboundedness assumption gives the following complete modified
log-Sobolev inequality.

\begin{corollary}Replacing the conditional hyperboundedness assumption in
Theorem~\ref{thm:kms-cmlsi} by the CB hyperboundedness assumption
\begin{equation}\label{eq:complete-hb}
 \|T_{t_0}:L_2(\sigma)
 \longrightarrow L_2^q(\mathcal N,\sigma)\|_{cb}\le M,
\end{equation}
we obtain the complete modified log-Sobolev inequality: for every reference system $\mathcal H_R$ and joint state $\rho$ on
$\mathcal H\otimes\mathcal H_R$,
\begin{equation}\label{eq:complete-mlsi-decay}
 D((T_{t*}\otimes\id_R)\rho\|(E_*\otimes\id_R)\rho)
 \le e^{-2\alpha t}D(\rho\|(E_*\otimes\id_R)\rho),\qquad t\ge 0
\end{equation}
with the same constant $\alpha$ as in \eqref{eq:kms-cmlsi-rate}.
\end{corollary}

Our proofs use a finite-step entropy comparison for quantum channels, whose small-time limit yields a
universal lower bound on entropy production in terms of square-root
energy. This comparison provides the regularity estimate needed in the
semigroup argument. An interesting ingredient is the use of the square-root remainder
\[
 W_\rho=(\sqrt\rho-\sqrt{E_*\rho})^2,
\]
whose trace measures the departure of $\rho$ from its equilibrium
reference $E_*\rho$. This remainder allows us to combine
hyperboundedness with the gap estimate and remove the defect
directly at the entropy level.

Section~\ref{sec:prelim} introduces invariant states and conditional
norms. Section~\ref{sec:entropy} develops entropy comparisons and
square-root estimates. Section~\ref{sec:tail} proves
Theorem~\ref{thm:main}, and Section~\ref{sec:semigroup-general}
establishes the semigroup criteria and proves
Theorem~\ref{thm:kms-cmlsi}.

\section{Conditional expectations and norms}\label{sec:prelim}
We first review the structure of the invariant states and the conditional
norms used in the hyperboundedness assumption. Set $d=\dim\mathcal H$
and let $\tau_d=\Tr/d$ denote the normalized trace on $B(\mathcal H)$.
A faithful conditional expectation $E:B(\mathcal H)\to\mathcal N$ is a
unital completely positive projection $E\circ E=E$ onto a unital $*$-subalgebra
$\mathcal N$, which satisfies the bimodule property and faithfulness
\begin{align*}
 & E(AXB)=AE(X)B, \qquad A,B\in\mathcal N, \qquad \text{ and }
 & E(X^*X)=0\Longrightarrow X=0.
\end{align*}
Its trace predual is denoted by $E_*$, and the same convention is used
for $\Phi_*$, where $\Tr\Phi_*(\rho)X=\Tr\rho\Phi(X)$.

\subsection{Invariant states and GNS contraction}
Recall the distinguished invariant state is $\omega_0=E_*(I/d)$.
The faithfulness of $E$ implies that $\omega_0>0$.
The predual $E_*$ is an $\mathcal N$-bimodule map; applying this
property to $I/d$ shows that $\omega_0\in\mathcal N'$.
For $a\in\mathcal N$, trace duality gives $\Tr\omega_0 a=\tau_d(a)$.

Let $E_*\sigma=\sigma$ be an invariant positive matrix. The restriction of
$\Tr\sigma(\,\cdot\,)$ to $\mathcal N$ has a unique density
$a\in\mathcal N_+$ with respect to the trace $\tau_d$ on $\mathcal N$. Moreover, 
\begin{equation}\label{eq:invariant-density}
 \sigma=\omega_0 a,\qquad a\in\mathcal N_+.
\end{equation}
Indeed, for arbitrary $X$,
\[
 \Tr\sigma X
 =\Tr\sigma E(X)
 =\tau_d\bigl(aE(X)\bigr)=\tau_d\bigl(E(X) a\bigr )
 =\Tr\omega_0 aX=\Tr a \omega_0 X .
\]
This also implies that $a$ and $\omega_0$ commute, and
$\supp\sigma=\supp a\in\mathcal N$.
The matrix $a$ itself need not be central in $\mathcal N$, and the density $\sigma$ need
not belong to $\mathcal N$.

For positive matrices $A,B$ of equal trace, we use
$D(A\|B)=\Tr A(\log A-\log B)$, with the usual support convention
and $D(0\|0)=0$. This extension is homogeneous:
$D(cA\|cB)=cD(A\|B)$ for $c>0$.
Since $E_*$ preserves
the trace, this convention applies to $D(A\|E_*A)$ for $A\ge0$.
Moreover,
\begin{equation}\label{eq:kms-homogeneous-pythagoras}
 \begin{aligned}
 D(A\|E_*A)
=&\Tr A(\log A-\log\omega_0)-\Tr E_*A(\log E_*A-\log\omega_0).
 \end{aligned}
\end{equation}

If a UCP map $\Phi$ preserves a faithful state $\sigma$ and fixes
$\mathcal N$ pointwise, then the $\sigma$-preserving conditional
expectation onto $\mathcal N$ satisfies $\Phi E=E\Phi=E$.
Indeed, the multiplicative-domain property makes $\Phi$ an
$\mathcal N$-bimodule map, and, for $B\in\mathcal N$,
\[
 \Tr\sigma B^*\Phi(X)=\Tr\sigma\Phi(B^*X)
 =\Tr\sigma B^*X.
\]
GNS orthogonality of $E$ gives $E\Phi=E$, while $\Phi E=E$
follows from fixedness. The same argument applies to each map of a QMS.

The compatibility relation $\Phi E=E\Phi=E$ implies that $\Phi$
fixes $\mathcal N$ pointwise and $\mathcal N$ therefore
belongs to the multiplicative domain of $\Phi$. Consequently,
$\Phi$ is also an $\mathcal N$-bimodule map
\[ \Phi(AXB)=A\Phi(X)B, \qquad A,B\in\mathcal N \]
Taking trace preduals shows
that $\Phi_*$ is also an $\mathcal N$-bimodule map and that
\begin{equation}\label{eq:predual-compatibility}
 E_*\Phi_*=\Phi_*E_*=E_*.
\end{equation}
In particular, $\Phi_*\sigma=\sigma$ whenever $E_*\sigma=\sigma$.

For a positive semi-definite matrix $\sigma$, define the GNS seminorm
$$\|X\|_{2,\mathrm{GNS},\sigma}^2=\Tr\sigma X^*X. $$
The contraction bound at $\omega_0$ extends to all invariant reference
states, including singular states.  

\begin{lemma}\label{lem:gns-transfer}
Let $E$ be a faithful conditional expectation and let $\Phi$ be
unital and completely positive with $\Phi E=E\Phi=E$.
Suppose that, for some $0\le s<1$,
\begin{equation}
 \|(\Phi-E)X\|_{2,\mathrm{GNS},\omega_0}
 \le s\|X\|_{2,\mathrm{GNS},\omega_0}
 \qquad(X\in B(\mathcal H))
\end{equation}
holds. Then for every positive semi-definite $\sigma$
satisfying $E_*\sigma=\sigma$, and every $X\in B(\mathcal H)$,
\begin{equation}\label{eq:gns-at-sigma}
 \|(\Phi-E)X\|_{2,\mathrm{GNS},\sigma}
 \le s\|X\|_{2,\mathrm{GNS},\sigma}.
\end{equation}
If $\sigma>0$, the estimates at $\sigma$ and $\omega_0$ are equivalent
with the same $s$.
\end{lemma}

\begin{proof}The proof follows from
\cite[Proposition~2.2]{GG} for faithful invariant states.
Write $\sigma=\omega_0 a$ as in \eqref{eq:invariant-density}.
Bimodularity gives
\[
 (\Phi-E)(Xa^{1/2})
 =((\Phi-E)X)a^{1/2}.
\]
Since $a$ commutes with $\omega_0$, we conclude that
\begin{align*}
 \|(\Phi-E)X\|_{2,\mathrm{GNS},\sigma}
 =\|(\Phi-E)(Xa^{1/2})\|_{2,\mathrm{GNS},\omega_0}
 \le s\|Xa^{1/2}\|_{2,\mathrm{GNS},\omega_0}
 =s\|X\|_{2,\mathrm{GNS},\sigma}.
\end{align*}
The argument uses no inverse of $a$ and applies when $\sigma$ is
singular. If $\sigma>0$, the map $X\mapsto Xa^{1/2}$ is bijective,
so the same identities give the converse.
\end{proof}

For a compatible QMS generator $\mathcal L$, bimodularity also gives
$\mathcal L(Xa^{1/2})=\mathcal L(X)a^{1/2}$.
If $\sigma>0$, the change of variables $Y=Xa^{1/2}$ is a bijection
of $\ker E$ and, for $X\ne0$,
\begin{equation}\label{eq:gns-gap-change-weight}
 \frac{-\Re\Tr\sigma X^*\mathcal L(X)}
      {\|X\|_{2,\mathrm{GNS},\sigma}^2}
 =\frac{-\Re\Tr\omega_0Y^*\mathcal L(Y)}
       {\|Y\|_{2,\mathrm{GNS},\omega_0}^2}.
\end{equation}
Thus the optimal GNS coercivity constant on $\ker E$ is the same
for the two references.

\subsection{Conditional hyperboundedness}
For $1\le p<\infty$ and a faithful state $\omega$, define the Kosaki weighted $L_p$-norm
\begin{equation}\label{eq:weighted-lp}
 \|X\|_{p,\omega}
 =\|\omega^{\frac{1}{2p}}X\omega^{\frac{1}{2p}}\|_{p},
\end{equation}
where $\|\cdot\|_{p}$ is the Schatten norm for the ordinary trace. We write $L_p(\omega)$ for the space equipped with this weighted norm.
For $X\in B(\mathbb C^n)$, the norm associated with the normalized trace is
\[
 \|X\|_{p,\tau_n}
 :=\bigl(\tau_n(|X|^p)\bigr)^{1/p}
 =n^{-1/p}\|X\|_{p}
 \qquad(1\le p<\infty).
\]
For $p=\infty$, both $\|\cdot\|_{p,\omega}$ and
$\|\cdot\|_{p,\tau_n}$ denote the operator norm. In particular, the $2$-norm is the KMS norm, which differs from the GNS norm,
\[
 \|X\|_{2,\omega}^2
 =\Tr\omega^{1/2}X^*\omega^{1/2}X,
 \qquad
 \|X\|_{2,\mathrm{GNS},\omega}^2
 =\Tr\omega X^*X.
\]
For the distinguished state
$\omega_0\in \mathcal N'$, 
\begin{equation}\label{eq:fixed-algebra-norms}
 \|A\|_{p,\omega_0}=\|A\|_{p,\tau_d}, \qquad 
 A\in\mathcal N, 1\le p\le\infty.
\end{equation}

Following
\cite[Section~3.4]{GJL}, we write $L_p^q(\mathcal N\subset B(\mathcal H),\omega_0)$ for the conditional norms below: for $1\le p,q\le\infty$ with $p\le q$, define
\begin{equation}\label{eq:conditional-inf}
 \|X\|_{L_p^q(\mathcal N\subset B(\mathcal H),\omega_0)}
 =\inf_{\substack{X=AZB\\A,B\in\mathcal N}}
 \|A\|_{2r,\omega_0}\|Z\|_{q,\omega_0}\|B\|_{2r,\omega_0},
 \qquad \frac1r=\frac1p-\frac1q,
\end{equation}
where the infimum runs over all factorizations $X=AZB$ with $A,B\in \mathcal{N}, Z\in B(\mathcal H)$; for $p\ge q$, put
\begin{equation}\label{eq:conditional-sup}
 \|X\|_{L_p^q(\mathcal N\subset B(\mathcal H),\omega_0)}
 =\sup_{A,B\in\mathcal N\setminus\{0\}}
 \frac{\|AXB\|_{q,\omega_0}}
 {\|A\|_{2r,\omega_0}\|B\|_{2r,\omega_0}},
 \qquad \frac1r=\frac1q-\frac1p.
\end{equation}
When the ambient algebra is clear, we abbreviate this notation to
$L_p^q(\mathcal N,\omega_0)$.
When $p=q$, both conditional norms coincide with the usual $p$-norm $\|X\|_{p,\omega_0}$.  We use the
same definitions with $\tau_d$ in place of $\omega_0$ for the
tracial conditional norms. 

For a faithful state $\sigma=E_*\sigma$, the standard weighted
amalgamated space in sandwiched coordinates
\cite[Section~1.1 and Chapter~2]{JP} has the following factorization.
For $2\le q\le\infty$ and $1/r=1/2-1/q$,
\begin{equation}\label{eq:conditional-weighted-inf}
 \|X\|_{L_2^q(\mathcal N,\sigma)}
 =\inf_{\substack{X=RZS\\R,S\in\mathcal N}}
 \|\sigma^{1/4}R\sigma^{-1/(2q)}\|_{2r}
 \|Z\|_{q,\sigma}
 \|\sigma^{-1/(2q)}S\sigma^{1/4}\|_{2r},
\end{equation}
where $Z\in B(\mathcal H)$ and $1/\infty=0$.
At $q=2$ this is the usual $L_2(\sigma)$ norm.
Writing $\sigma=\omega_0a$ as in~\eqref{eq:invariant-density},
the bijective substitutions
\[
 R'=a^{1/4}Ra^{-1/(2q)},\quad
 Z'=a^{1/(2q)}Za^{1/(2q)},\quad
 S'=a^{-1/(2q)}Sa^{1/4}
\]
in~\eqref{eq:conditional-weighted-inf} give the change-of-weight identity
\begin{equation}\label{eq:conditional-change-weight}
 \|X\|_{L_2^q(\mathcal N,\sigma)}
 =\|a^{1/4}Xa^{1/4}\|_{L_2^q(\mathcal N,\omega_0)}.
\end{equation}
Here $[\omega_0,\mathcal N]=0$ identifies the three factor norms
with their counterparts in~\eqref{eq:conditional-inf}.
The map $X\mapsto a^{1/4}Xa^{1/4}$ is also an isometry from
$L_2(\sigma)$ onto $L_2(\omega_0)$ and commutes with every
$\mathcal N$-bimodule map $\Psi$. Consequently,
\begin{equation}\label{eq:hb-change-weight}
 \|\Psi:L_2(\sigma)\to L_2^q(\mathcal N,\sigma)\|
 =\|\Psi:L_2(\omega_0)\to L_2^q(\mathcal N,\omega_0)\|.
\end{equation}
The same identity holds at every auxiliary level, with references
$\sigma\otimes I_m/m$, $\omega_0\otimes I_m/m$ and density
$a\otimes I_m$, so uniform amplification bounds retain the same constant.

For $2\le q\le\infty$, the conditional norm dominates the KMS $L_2$ norm
\cite[Proposition~3.1(iii)]{BR}:
\[
 \|Y\|_{2,\omega_0}
 \le\|Y\|_{L_2^q(\mathcal N,\omega_0)}
 \qquad(Y\in B(\mathcal H)).
\]
Since $\Phi(I_d)=I_d$ and the identity has norm one in both spaces,
the hyperboundedness constant satisfies $M\ge1$.

\subsection{Square-root geometry and KMS contraction}
For $\sigma>0$, $t\in\mathbb R$, and a UCP map $\Phi$, define
\begin{equation}\label{eq:static-kms-implementation}
 \Gamma_\sigma^t(X)=\sigma^{t/2}X\sigma^{t/2},\qquad
 \widehat{\Phi}_\sigma^{\,*}=\Gamma_\sigma^{-1/2}\Phi_*\Gamma_\sigma^{1/2}.
\end{equation}
We denote by 
$\langle X,Y\rangle_{\mathrm{HS}}=\Tr X^*Y$ the Hilbert--Schmidt inner product and by $\|\cdot\|_2$ its norm.
The following projection estimate depends only on $E$.
\begin{lemma}\label{lem:canonical-root-projection}
Let $E$ be a faithful conditional expectation and $\omega_0=E_*(I/d)$ be the distinguished invariant state.
Then $P=\Gamma_{\omega_0}^{1/2}E\Gamma_{\omega_0}^{-1/2}$ is completely positive and a Hilbert--Schmidt  orthogonal projection. For every self-adjoint $A$,
\begin{equation}\label{eq:kms-projection-square}
 (PA)^2\le E_*(A^2).
\end{equation}
Consequently, every density operator $\rho$ satisfies
\begin{equation}\label{eq:projection-root-mass}
  \Tr(\sqrt\rho-\sqrt{E_*\rho})^2
 \le2\|(I-P)\sqrt\rho\|_2^2.
\end{equation}
\end{lemma}
\begin{proof}
The expectation $E$ is orthogonal on $L_2(\omega_0)$, and
$\Gamma_{\omega_0}^{1/2}$ is an isometry from $L_2(\omega_0)$ onto Hilbert--Schmidt space
\cite[Eqs.~(2.13)--(2.15)]{BR}. Hence $P$ is a
Hilbert--Schmidt orthogonal projection; its complete positivity
follows from that of $E$ and $\Gamma_{\omega_0}^{\pm1/2}$.

Since $\omega_0\in\mathcal N'$, $P$ is an $\mathcal N$-bimodule map
with $\operatorname{Ran}P=\sqrt{\omega_0}\mathcal N$.
For $A=A^*$, we have $(PA)^2\in\omega_0\mathcal N$, so
$E_*((PA)^2)=(PA)^2$ by bimodularity and $E_*\omega_0=\omega_0$.
For every $B\ge0$, $E(B)^{1/2}\in\mathcal N$, and hence
\begin{align*}
 \Tr B(PA)^2=\Tr E(B)(PA)^2=\bigl\|P\bigl(AE(B)^{1/2}\bigr)\bigr\|_2^2
 \le\bigl\|AE(B)^{1/2}\bigr\|_2^2=\Tr B\,E_*(A^2).
\end{align*}
Since $B\ge0$ is arbitrary, this proves~\eqref{eq:kms-projection-square}.

For a density matrix $\rho$, applying~\eqref{eq:kms-projection-square}
to $\sqrt\rho$ gives $0\le P(\sqrt\rho)\le\sqrt{E_*\rho}$. Also,
$\sqrt{E_*\rho}\in\sqrt{\omega_0}\mathcal N=\operatorname{Ran}P$. Hence
\begin{align*}
 & \Tr(\sqrt\rho-\sqrt{E_*\rho})^2=2-2\Tr P(\sqrt\rho)\sqrt{E_*\rho}\le2-2\Tr\bigl[P(\sqrt\rho)^2\bigr]=2\|(I-P)\sqrt\rho\|_2^2. \hfill \qedhere
\end{align*}
\end{proof}

\begin{lemma}\label{lem:static-kms-contraction}
Let $\Phi$ be unital and completely positive, and let $\sigma>0$
be a faithful invariant state satisfying
$\Phi_*\sigma=\sigma$. Then $\widehat{\Phi}_\sigma^{\,*}$ is a
Hilbert--Schmidt contraction.
Let $E$ be a faithful conditional expectation with $\Phi E=E\Phi=E$.
Suppose that, for some $0\le s<1$,
\[
 \|(\Phi-E)X\|_{2,\mathrm{GNS},\omega_0}
 \le s\|X\|_{2,\mathrm{GNS},\omega_0}
 \qquad X\in B(\mathcal H).
\]
Then
\begin{equation}\label{eq:static-kms-contraction}
 \|\widehat{\Phi}_{\omega_0}^{\,*}-P\|_{2\to2}
 =\|\Gamma_{\omega_0}^{1/2}(\Phi-E)\Gamma_{\omega_0}^{-1/2}\|_{2\to2}\le s.
\end{equation}
Independently of the GNS bound, the relation $\Phi E=E\Phi=E$ yields
\begin{equation}\label{eq:invariant-kms-implementation}
 P\widehat{\Phi}_{\omega_0}^{\,*}=\widehat{\Phi}_{\omega_0}^{\,*}P=P,
 \qquad \widehat{\Phi}_\sigma^{\,*}=\widehat{\Phi}_{\omega_0}^{\,*},
\end{equation}
where the second identity holds provided $E_*\sigma=\sigma$.
In particular, KMS symmetry with respect to such a $\sigma$ is equivalent
to KMS symmetry with respect to $\omega_0$.
\end{lemma}

\begin{proof}
A $\sigma$-preserving channel is contractive on the KMS space
$L_2(\sigma)$ \cite[Corollary~3.3]{Wirth26}. Thus, for every $Y\in B(\mathcal H)$,
\begin{align*}
 \|\Gamma_\sigma^{1/2}\Phi\Gamma_\sigma^{-1/2}(Y)\|_2
 =\|\Phi(\Gamma_\sigma^{-1/2}(Y))\|_{2,\sigma}\le\|\Gamma_\sigma^{-1/2}(Y)\|_{2,\sigma}=\|Y\|_2.
\end{align*}
Since
\[
 \bigl(\Gamma_\sigma^{1/2}\Phi\Gamma_\sigma^{-1/2}\bigr)^*
 =\Gamma_\sigma^{-1/2}\Phi_*\Gamma_\sigma^{1/2}=\widehat{\Phi}_\sigma^{\,*},
\]
both $\Gamma_\sigma^{1/2}\Phi\Gamma_\sigma^{-1/2}$ and $\widehat{\Phi}_\sigma^{\,*}$
are Hilbert--Schmidt contractions.

By the GNS--KMS contraction comparison
\cite[Theorem~3.1]{Wirth26}, the assumed GNS estimate gives
\[
 \|(\Phi-E)X\|_{2,\omega_0}
 \le s\|X\|_{2,\omega_0}
 \qquad(X\in B(\mathcal H)).
\]
Conjugating by $\Gamma_{\omega_0}^{1/2}$ and taking
Hilbert--Schmidt adjoints proves~\eqref{eq:static-kms-contraction},
since $P=P^*$.

Finally, $\Phi E=E\Phi=E$ gives $P\widehat{\Phi}_{\omega_0}^{\,*}=\widehat{\Phi}_{\omega_0}^{\,*}P=P$.
If $E_*\sigma=\sigma$, write $\sigma=\omega_0a$ as
in~\eqref{eq:invariant-density}.
The matrix $a>0$ belongs to $\mathcal N$ and commutes with $\omega_0$.
Bimodularity gives
\begin{align*}
&\widehat{\Phi}_\sigma^{\,*}=\Gamma_{\sigma}^{-1/2}\Phi_*\Gamma_{\sigma}^{1/2}=\Gamma_{\omega_0}^{-1/2}\Gamma_a^{-1/2}\Phi_*\Gamma_a^{1/2}\Gamma_{\omega_0}^{1/2}=\Gamma_{\omega_0}^{-1/2}\Phi_*\Gamma_{\omega_0}^{1/2}
 =\widehat{\Phi}_{\omega_0}^{\,*}.
\end{align*}
Taking Hilbert--Schmidt adjoints gives the same identity for the
Heisenberg implementations, proving the assertion about KMS symmetry.
\end{proof}

\section{Entropy comparisons and square-root remainders}\label{sec:entropy}
In this section, we establish a finite-step entropy comparison for channels with a faithful invariant state, whose small-time limit gives
the energy comparisons in Section~\ref{sec:semigroup-general}. We also obtain a static conditional entropy estimate
for the remainder that will be used in the semigroup argument.

\subsection{Finite-step entropy comparison}

\begin{proposition}\label{prop:finite-step-comparison}
Let $\Phi$ be unital and completely positive, with a faithful
invariant density matrix $\sigma=\Phi_*(\sigma)$. For faithful density matrices
$\rho,\tau$, define 
\[
a_{\tau,\rho}:=\langle\sqrt\tau,\widehat{\Phi}_\sigma^{\,*}\sqrt\rho\rangle_{\mathrm{HS}}, \qquad a_{\rho}:=a_{\rho,\rho}.
\]
Then $0\le a_{\tau,\rho}\le1$ and
\begin{equation}\label{eq:finite-step-comparison}
 D(\rho\|\sigma)-D(\Phi_*\rho\|\sigma)+D(\Phi_*\rho\|\tau)
 \ge\pi\,\frac{1-a_{\tau,\rho}}{1+a_{\tau,\rho}}.
\end{equation}
In particular, for every state $\rho$,
\begin{equation}\label{eq:finite-step-norm-loss}
 D(\rho\|\sigma)-D(\Phi_*\rho\|\sigma)
 \ge\frac\pi4\bigl(1-\|\widehat{\Phi}_\sigma^{\,*}\sqrt\rho\|_2^2\bigr).
\end{equation}
If $\Phi$ is KMS-symmetric with respect to $\sigma$ and
$\tau=\rho$, then
\begin{equation}\label{eq:finite-step-kms-comparison}
 D(\rho\|\sigma)-D(\Phi_*\rho\|\sigma)+D(\Phi_*\rho\|\rho)
 \ge2\pi\,\frac{1-a_\rho}{1+a_\rho}.
\end{equation}
\end{proposition}

\begin{proof}
For $0\le\Re z\le1$, define
\[
 F(z)=\sigma^{-z/2}\Phi_*
       (\sigma^{z/2}\rho^{1-z}\sigma^{z/2})\sigma^{-z/2},
 \qquad f(z)=\Tr\tau^zF(z).
\]
These functions are analytic in $z$ and bounded on the closed strip.
On its left boundary, unitary invariance and the trace-norm
contraction of the CPTP map $\Phi_*$ give $\|F(iy)\|_1\le1$.
Since $\|\tau^{iy}\|_\infty=1$, we have $|f(iy)|\le1$.
For the right boundary, invariance of $\sigma$ makes
$\Gamma_\sigma^{-1}\Phi_*\Gamma_\sigma$ unital and completely
positive. Hence
\[
 \|F(1+iy)\|_\infty
 =\bigl\|(\Gamma_\sigma^{-1}\Phi_*\Gamma_\sigma)
       (\sigma^{iy/2}\rho^{-iy}\sigma^{iy/2})\bigr\|_\infty\le1.
\]
Together with $\|\tau^{1+iy}\|_1=1$, this yields
$|f(1+iy)|\le1$. The Hadamard three-lines theorem therefore gives
$|f(z)|\le1$ throughout the strip.
Trace preservation, invariance, and direct differentiation give
\begin{equation}\label{eq:comparison-analytic-data}
 \begin{aligned}
 f(0)&=f(1)=1,\qquad f(1/2)=a_{\tau,\rho},\\
 -f'(0)&=D(\rho\|\sigma)-D(\Phi_*\rho\|\sigma)+D(\Phi_*\rho\|\tau).
 \end{aligned}
\end{equation}
The strip bound and the midpoint identity in~\eqref{eq:comparison-analytic-data},
together with positivity of $\widehat{\Phi}_\sigma^{\,*}$, give $0\le a_{\tau,\rho}\le1$.

If $f\equiv1$, the two-state comparisons are immediate. Otherwise the
maximum modulus principle gives $|f|<1$ in the open strip, so
\[
 g(z)=\frac{1-f(z)}{1+f(z)}
\]
is analytic there, with $\Re g\ge0$. It is also analytic near
$z=0$, where
\[
 g(0)=0,\qquad
 g'(0)=-\frac12f'(0),\qquad
 g(1/2)=\frac{1-a_{\tau,\rho}}{1+a_{\tau,\rho}}.
\]
The conformal map
\[
 \varphi(z)=\tan\!\left(\frac\pi2(z-1/2)\right)
\]
sends the open strip onto the unit disk, with
$\varphi(0)=-1$ and $\varphi(1/2)=0$.
Set $\widetilde g=g\circ\varphi^{-1}$.
Harnack's inequality for the positive harmonic function $\Re\widetilde g$ gives
\begin{equation}\label{eq:comparison-harnack}
 \Re g(x)\ge
 \frac{1+\varphi(x)}{1-\varphi(x)}\Re g(1/2)
 =\tan\frac{\pi x}{2}\,\Re g(1/2)
 \qquad(0<x\le1/2).
\end{equation}
Divide by $x$ and let $x\downarrow0$.
Equation~\eqref{eq:comparison-analytic-data} then gives
\eqref{eq:finite-step-comparison}.

For~\eqref{eq:finite-step-norm-loss}, first let $\rho$ be faithful.
Take $\tau=\Phi_*\rho$ in~\eqref{eq:finite-step-comparison}. Cauchy--Schwarz gives
\[
 a_{\Phi_*\rho,\rho }=\langle\sqrt\tau,\widehat{\Phi}_\sigma^{\,*}\sqrt\rho\rangle_{\mathrm{HS}}\le\|\sqrt{\Phi_*\rho}\|_2\|\widehat{\Phi}_\sigma^{\,*}\sqrt\rho\|_2
 =\|\widehat{\Phi}_\sigma^{\,*}\sqrt\rho\|_2.
\]
Consequently,
\begin{align*}
 D(\rho\|\sigma)-D(\Phi_*\rho\|\sigma)
 \ge\pi\frac{1-a_{\Phi_*\rho,\rho }}{1+a_{\Phi_*\rho,\rho }}
 \ge\frac\pi4(1-a_{\Phi_*\rho,\rho }^2)
 \ge\frac\pi4\bigl(1-\|\widehat{\Phi}_\sigma^{\,*}\sqrt\rho\|_2^2\bigr).
\end{align*}

Now suppose that $\Phi$ is KMS-symmetric and $\tau=\rho$.
For $z\in\mathbb C$, set
\[
 V_z=\sigma^{z/2-1/4}\rho^{1-z}\sigma^{z/2-1/4}.
\]
Since $\widehat{\Phi}_\sigma^{\,*}$ is Hilbert--Schmidt self-adjoint and preserves adjoints,
\[
 f(z)=\Tr V_{1-z}\widehat{\Phi}_\sigma^{\,*}(V_z)
     =\Tr V_z\widehat{\Phi}_\sigma^{\,*}(V_{1-z})=f(1-z).
\]
Consequently,
\begin{equation}\label{eq:comparison-reflection}
 g(1-z)=g(z)
 \qquad(0<\Re z<1).
\end{equation}
Since $\varphi(1-z)=-\varphi(z)$, the disk function
$\widetilde g$ is even. Its Taylor series defines an analytic
function $H$ on the disk such that $\widetilde g(w)=H(w^2)$.
Every point of the disk has a square root in the disk, so
$\Re H\ge0$. Harnack's inequality applied to $\Re H$ yields
\begin{equation}\label{eq:comparison-kms-harnack}
 \Re g(x)\ge
 \frac{1-\varphi(x)^2}{1+\varphi(x)^2}\Re g(1/2)
 =\sin(\pi x)\,\Re g(1/2)
 \qquad(0<x\le1/2).
\end{equation}
Dividing by $x$ and letting $x\downarrow0$, we obtain
\eqref{eq:finite-step-kms-comparison} from
\eqref{eq:comparison-analytic-data} with $\tau=\rho$.
\end{proof}

\subsection{Midpoint tails and dissipation budgets}
For positive matrices $A,B$ and $t>0$, define 
\begin{equation}\label{eq:hockey-stick}
 e_t(A\mid B)=\Tr(A-tB)_+
 =\sup_{0\le K\le I}\Tr K(A-tB),
\end{equation}
where $X_+$ denotes the positive part of a self-adjoint matrix $X$.
For $L\ge 1$, 
\begin{equation}\label{eq:tail-functionals}
 F_L(A\mid B)=\int_L^\infty e_t(A\mid B)\,\frac{dt}{t}
,\qquad
 F(A\mid B)=F_1(A\mid B).
\end{equation}
The integral representation of quantum relative entropy
\cite[Theorem~6]{Frenkel}, in the positive-part form of
\cite[Corollary~2.3]{HT}, states that for states $A,B$,
\begin{equation}\label{eq:hockey-stick-entropy}
 D(A\|B)=\int_1^\infty
 \left[\frac{e_t(A\mid B)}t+\frac{e_t(B\mid A)}{t^2}\right]dt.
\end{equation}

We bound the part of $D(A\|B)$ outside its high tail by a reverse
entropy at the midpoint.
\begin{lemma}\label{lem:midpoint-comparison}
Let $\rho,\sigma$ be states and put
$K_0=(2\log2-1)^{-1}$. Then
\begin{equation}\label{eq:entropy-midpoint-tail}
 D(\rho\|\sigma)\le2K_0D\left(\sigma\middle\|\frac{\rho+\sigma}{2}\right)
 +F_4(\rho\mid\sigma).
\end{equation}
The coefficient $2K_0$ is optimal, even for commuting states.
\end{lemma}

\begin{proof}
The midpoint case of the skew-divergence representation
\cite[Proposition~2.9]{HT} gives
\begin{equation}\label{eq:midpoint-tail-kernel}
 D\left(\sigma\middle\|\frac{\rho+\sigma}{2}\right)=\int_1^\infty
 \left[\frac{e_t(\rho\mid\sigma)}{(1+t)^2}
       +\frac{e_t(\sigma\mid\rho)}{t(1+t)^2}\right]dt.
\end{equation}
Recall Chebyshev's inequality: for a probability measure $\mu$ on an interval $[a,b]$ and bounded nonincreasing functions
$f,g$, we have 
\begin{equation}\label{eq:monotone-integral}
 \int fg\,d\mu\ge
 \left(\int f\,d\mu\right)\left(\int g\,d\mu\right).
\end{equation}
Note that the functions $e_t$ in~\eqref{eq:hockey-stick} are bounded and nonincreasing.
Thus, for $L>1$, 
\begin{align*}
 \int_1^L\frac{e_t(\rho\mid\sigma)}t\,dt
 &\le\frac{2(L+1)}{L-1}\log L
      \int_1^L\frac{e_t(\rho\mid\sigma)}{(1+t)^2}\,dt,\\
 \int_1^\infty\frac{e_t(\sigma\mid\rho)}{t^2}\,dt
 &\le2K_0\int_1^\infty
          \frac{e_t(\sigma\mid\rho)}{t(1+t)^2}\,dt.
\end{align*}
Splitting~\eqref{eq:hockey-stick-entropy} at $L$ and
using~\eqref{eq:midpoint-tail-kernel} gives
\[
 D(\rho\|\sigma)\le
 \max\left\{2K_0,\frac{2(L+1)}{L-1}\log L\right\}
 D\left(\sigma\middle\|\frac{\rho+\sigma}{2}\right)
 +F_L(\rho\mid\sigma).
\]
Since
$(10/3)\log4<2K_0<5\log3$, the choice $L=4$
proves~\eqref{eq:entropy-midpoint-tail}, while $L=9$ gives
\begin{equation}\label{eq:midpoint-nine-tail}
 D(\rho\|\sigma)\le5\log3\,
 D\left(\sigma\middle\|\frac{\rho+\sigma}{2}\right)+F_9(\rho\mid\sigma).
\end{equation}
To prove optimality, take
\[
 \rho=\operatorname{diag}(0,1),\qquad
 \sigma_\epsilon=\operatorname{diag}(\epsilon,1-\epsilon).
\]
For $\epsilon\in (0,1)$, 
\begin{align*}
 D\left(\sigma_\epsilon\middle\|\frac{\rho+\sigma_\epsilon}{2}\right)
 =\left(\log2-\frac12\right)\epsilon+O(\epsilon^2),\qquad 
 D(\rho\|\sigma_\epsilon)
 =\epsilon+O(\epsilon^2).
\end{align*}
For $0<\epsilon<3/4$, we have
$\rho\le4\sigma_\epsilon$ and therefore $F_4(\rho\mid\sigma_\epsilon)=0$.
The two expansions give the limiting ratio $2K_0$ in
\eqref{eq:entropy-midpoint-tail}.
\end{proof}

For a state $\rho$, we write $\sigma=E_*\rho$ throughout
the remainder of this section and define the positive matrix
\begin{equation}\label{eq:sqrt-remainder}
 W_\rho=(\sqrt\rho-\sqrt{E_*\rho})^2.
\end{equation}
Its trace $\Tr W_\rho=\|\sqrt\rho-\sqrt{E_*\rho}\|_2^2$ is the
squared quantum Hellinger distance in the normalization of
\cite[Definition~2.8]{OW}.

The GNS midpoint estimate of \cite[Lemma~3.2 and Eq.~(3.7)]{GG}
controls the output midpoint entropy. The finite-step comparison
controls $\Tr W_\rho$ by the same entropy loss.

\begin{lemma}\label{lem:entropy-budget}
Let $E$ be a faithful conditional expectation and let $\Phi$ be
unital and completely positive with $\Phi E=E\Phi=E$.
Suppose that, for some $0\le s<1$,
\[
 \|(\Phi-E)X\|_{2,\mathrm{GNS},\omega_0}
 \le s\|X\|_{2,\mathrm{GNS},\omega_0}
 \qquad(X\in B(\mathcal H)).
\]
Set $\kappa=1-s^2$, and for every state $\rho$, set
\[
 \Delta(\rho)=D(\rho\|E_*\rho)-D(\Phi_*\rho\|E_*\rho).
\]
Then
\begin{equation}\label{eq:midpoint-budget}
 D\left(E_*\rho\middle\|\frac{E_*\rho+\Phi_*\rho}{2}\right)
 \le\frac2\kappa \Delta(\rho),
\end{equation}
and
\begin{equation}\label{eq:hellinger-budget}
 \Tr W_\rho\le\frac8{\pi\kappa}\Delta(\rho).
\end{equation}
\end{lemma}

\begin{proof}
First consider a faithful state $\rho>0$. Compatibility gives
$\Phi_*E_*\rho=E_*\rho$, and $E_*\Phi_*\rho=E_*\rho$.
Set $c_\rho=(\rho+E_*\rho)/2$, so that
$\Phi_*c_\rho=(\Phi_*\rho+E_*\rho)/2$.
A direct calculation gives
\[
\begin{aligned}
&D(\rho\|c_\rho)+D(E_*\rho\|c_\rho)
-D(\Phi_*\rho\|\Phi_*c_\rho)-D(E_*\rho\|\Phi_*c_\rho)\\
&=\Tr\!\left(
\rho\log\rho-\Phi_*\rho\log\Phi_*\rho
-2c_\rho\log c_\rho
+2\Phi_*c_\rho\log\Phi_*c_\rho
\right)\\
&=D(\rho\|E_*\rho)-D(\Phi_*\rho\|E_*\rho)
-2D(c_\rho\|E_*\rho)+2D(\Phi_*c_\rho\|E_*\rho)\\
&=\Delta(\rho)-2\left[
D(c_\rho\|E_*\rho)-D(\Phi_*c_\rho\|E_*\rho)
\right].
\end{aligned}
\]
Data processing makes the left-hand side and the bracket on the right
nonnegative. Lemma~\ref{lem:gns-transfer} gives the same GNS contraction
coefficient $s$ at $E_*\rho$. Since $E_*c_\rho=E_*\rho$
and $c_\rho\ge E_*\rho/2$, \cite[Lemma~3.2]{GG} with $a=1/2$ yields
\[
 \frac{1}{2}\Delta(\rho)
 \ge D(c_\rho\|E_*\rho)-D(\Phi_*c_\rho\|E_*\rho)
 \ge\frac{\kappa}{4}D(E_*\rho\|\Phi_*c_\rho).
\]
This proves~\eqref{eq:midpoint-budget}.

Next, Lemma~\ref{lem:static-kms-contraction} and the orthogonality
of $P$ give
\[
 \begin{aligned}
 \|\widehat{\Phi}_{E_*\rho}^{\,*}\sqrt\rho\|_2^2
 \le\|P\sqrt\rho\|_2^2+s^2\|(I-P)\sqrt\rho\|_2^2=1-\kappa\|(I-P)\sqrt\rho\|_2^2.
 \end{aligned}
\]
Equations~\eqref{eq:finite-step-norm-loss} and
\eqref{eq:projection-root-mass} therefore yield
\[
 \Delta(\rho)\ge\frac\pi4\bigl(1-\|\widehat{\Phi}_{E_*\rho}^{\,*}\sqrt\rho\|_2^2\bigr)
 \ge\frac{\pi\kappa}{4}\|(I-P)\sqrt\rho\|_2^2
 \ge\frac{\pi\kappa}{8}\Tr W_\rho.
\]
The case of general $\rho$ follows by approximation with the faithful states 
$\rho_\epsilon=(1-\epsilon)\rho+\epsilon\omega_0$.
\end{proof}

\subsection{Square-root remainders and tail reduction}

The square-root remainder has a controlled conditional marginal.
A single Young-inequality argument also transfers the relevant tails
between the state and its remainder. This comparison will also give
a static entropy estimate for Section~\ref{sec:semigroup-general}.

\begin{lemma}\label{lem:sqrt-comparison}
Let $E$ be a faithful conditional expectation, and let $\Phi$ be unital and completely positive
with $\Phi E=E\Phi=E$.
Then for every state $\rho$, 
\begin{equation}\label{eq:sqrt-remainder-marginal}
 E_*W_\rho\le4E_*\rho,
\end{equation}
and
\begin{equation}\label{eq:adaptive-sqrt-comparison}
 F(W_\rho\mid E_*\rho)\le D(\rho\|E_*\rho)+(\log4)\Tr W_\rho,
 \qquad
 F_4(\Phi_*\rho\mid E_*\rho)\le F(\Phi_*W_\rho\mid E_*\rho).
\end{equation}
We also have
\begin{equation}\label{eq:static-sqrt-tail}
 F_4(\rho\mid E_*\rho)\le F(W_\rho\mid E_*\rho).
\end{equation}
\end{lemma}

\begin{proof}
For matrices $A,B$ and $a>0$, the operator Young inequality reads
\[
 (A+B)(A+B)^*
 \le(1+a^{-1})AA^*+(1+a)BB^*.
\]
Applying it to $(A,B)=(\sqrt\rho,-\sqrt\sigma)$ and
$(\sqrt\rho-\sqrt\sigma,\sqrt\sigma)$, respectively, gives, for $a\ge1$,
\[
 W_\rho\le(1+a^{-1})\rho+(1+a)\sigma,\qquad
 \rho\le(1+a^{-1})W_\rho+(1+a)\sigma.
\]
Taking $a=1$ in the first bound and applying $E_*$ gives
\eqref{eq:sqrt-remainder-marginal}.

We use the following tail transfer. If $X,Y,Z\ge0$ satisfy
$X\le(1+a^{-1})Y+(1+a)Z$ for every $a\ge1$, then
\begin{equation}\label{eq:young-tail-transfer}
 F_{(a_0+1)^2}(X\mid Z)\le F_{a_0^2}(Y\mid Z), \qquad a_0\ge 1
.
\end{equation}
Indeed,
\[
 X-(a+1)^2Z\le\frac{a+1}{a}(Y-a^2Z),
\]
so~\eqref{eq:hockey-stick} yields
\[
 e_{(a+1)^2}(X\mid Z)
 \le\frac{a+1}{a}e_{a^2}(Y\mid Z).
\]
With $t=(a+1)^2$ and then $u=a^2$, integration gives
\begin{align*}
 F_{(a_0+1)^2}(X\mid Z)=\int_{a_0}^\infty
       e_{(a+1)^2}(X\mid Z)\,\frac{2\,da}{a+1}\le\int_{a_0}^\infty e_{a^2}(Y\mid Z)\,\frac{2\,da}{a}=F_{a_0^2}(Y\mid Z).
\end{align*}
For the remaining applications in this proof, take $a_0=1$.

Apply~\eqref{eq:young-tail-transfer} with $(X,Y,Z)=(W_\rho,\rho,E_*\rho)$
to obtain $F_4(W_\rho\mid E_*\rho)\le F(\rho\mid E_*\rho)\le D(\rho\|E_*\rho)$.
The remaining part of $F(W_\rho\mid E_*\rho)$ is bounded by its mass:
\begin{align*}
 F(W_\rho\mid E_*\rho)
 =\int_1^4 e_t(W_\rho\mid E_*\rho)\,\frac{dt}{t}
   +F_4(W_\rho\mid E_*\rho)\le(\log4)\Tr W_\rho+D(\rho\|E_*\rho).
\end{align*}
This proves the first bound in~\eqref{eq:adaptive-sqrt-comparison}.

The second Young inequality and~\eqref{eq:young-tail-transfer} with
$(X,Y,Z)=(\rho,W_\rho,E_*\rho)$ give~\eqref{eq:static-sqrt-tail}.
Applying $\Phi_*$ to the same Young inequality, and using
$\Phi_*E_*\rho=E_*\rho$, gives
\[
 \Phi_*\rho\le(1+a^{-1})\Phi_*W_\rho+(1+a)E_*\rho.
\]
A final application of~\eqref{eq:young-tail-transfer} gives the second
bound in~\eqref{eq:adaptive-sqrt-comparison}.
\end{proof}

\begin{lemma}\label{lem:center-root-entropy}
Let $E$ be a faithful conditional expectation.
Every density matrix $\rho$ satisfies
\begin{equation}\label{eq:center-root-entropy}
 D(\rho\|E_*\rho)
 \le     5\log3\,D\left(E_*\rho\middle\|\frac{\rho+E_*\rho}{2}\right)+D(W_\rho\|E_*W_\rho).
\end{equation}
\end{lemma}

\begin{proof}
By~\eqref{eq:midpoint-nine-tail},
\[
 D(\rho\|E_*\rho)
 \le5\log3\,D\left(E_*\rho\middle\|\frac{\rho+E_*\rho}{2}\right)
       +F_9(\rho\mid E_*\rho).
\]
Using~\eqref{eq:young-tail-transfer} with $a_0=2$ and
$(X,Y,Z)=(\rho,W_\rho,E_*\rho)$, followed by
\eqref{eq:sqrt-remainder-marginal}, gives
\begin{align*}
 F_9(\rho\mid E_*\rho)
 \le F_4(W_\rho\mid E_*\rho)\le F(W_\rho\mid E_*W_\rho)
 \le D(W_\rho\|E_*W_\rho).
\end{align*}
Here the last inequality is the homogeneous form of
\eqref{eq:hockey-stick-entropy}, since $E_*$ preserves the trace.
\end{proof}

Lemmas~\ref{lem:entropy-budget} and~\ref{lem:sqrt-comparison},
together with~\eqref{eq:entropy-midpoint-tail}, reduce entropy
contraction to an estimate on the integrated tail of $W_\rho$.

\begin{proposition}\label{prop:closure}
Let $E$ be a faithful conditional expectation with the distinguished invariant state $\omega_0=E_*(I/d)$, and let $\Phi$ be
unital and completely positive with $\Phi E=E\Phi=E$.
Suppose that, for some $0\le s<1$,
\[
 \|(\Phi-E)X\|_{2,\mathrm{GNS},\omega_0}
 \le s\|X\|_{2,\mathrm{GNS},\omega_0},
 \qquad X\in B(\mathcal H).
\]
Set $\kappa=1-s^2$. Suppose that $0\le\theta<1$ and $B\ge0$ satisfy
\begin{equation}\label{eq:abstract-tail-contraction}
 F(\Phi_*W_\rho\mid E_*\rho)
 \le\theta F(W_\rho\mid E_*\rho)+B\Tr W_\rho
\end{equation}
for every state $\rho$, with $W_\rho$ as
in~\eqref{eq:sqrt-remainder}.
Then every state $\rho$ satisfies
\begin{equation}\label{eq:generic-closure}
 D(\Phi_*\rho\|E_*\rho)
 \le(1-\varepsilon)D(\rho\|E_*\rho),\qquad
 \varepsilon=
 \frac{\kappa(1-\theta)}
 {\kappa+4K_0+(8/\pi)(B+\theta\log4)}.
\end{equation}
\end{proposition}

\begin{proof}
Fix a state $\rho$ and recall that 
\[
 \Delta(\rho)=D(\rho\|E_*\rho)-D(\Phi_*\rho\|E_*\rho).
\]
By~\eqref{eq:entropy-midpoint-tail}, \eqref{eq:midpoint-budget}, and
the second inequality in~\eqref{eq:adaptive-sqrt-comparison},
\begin{align*}
 D(\rho\|E_*\rho)
 &=\Delta(\rho)+D(\Phi_*\rho\|E_*\rho)\\
 &\le\left(1+\frac{4K_0}{\kappa}\right)\Delta(\rho)
   +F(\Phi_*W_\rho\mid E_*\rho).
\end{align*}
By~\eqref{eq:abstract-tail-contraction} and the first inequality
in~\eqref{eq:adaptive-sqrt-comparison},
\begin{align*}
 F(\Phi_*W_\rho\mid E_*\rho)
 &\le\theta F(W_\rho\mid E_*\rho)+B\Tr W_\rho\\
 &\le\theta D(\rho\|E_*\rho)+(B+\theta\log4)\Tr W_\rho.
\end{align*}
Combining these estimates with~\eqref{eq:hellinger-budget} and
subtracting $\theta D(\rho\|E_*\rho)$, we obtain
\[
 (1-\theta)D(\rho\|E_*\rho)
 \le\frac{\kappa+4K_0+(8/\pi)(B+\theta\log4)}{\kappa}\Delta(\rho).
\]
By the value of $\varepsilon$ in~\eqref{eq:generic-closure}, this gives
\[
 D(\rho\|E_*\rho)-D(\Phi_*\rho\|E_*\rho)=\Delta(\rho)\ge\varepsilon D(\rho\|E_*\rho),
\]
which proves~\eqref{eq:generic-closure}.
\end{proof}

\section{Conditional hyperboundedness and entropy contraction}\label{sec:tail}

Proposition~\ref{prop:closure} reduces entropy contraction to an integrated
tail bound. We first derive a conditional second-moment estimate from
hyperboundedness, then construct a square-root truncation satisfying both
its order and conditional marginal hypotheses. The truncation error is
controlled by the input tail, which allows us to integrate the resulting
output-tail estimate. Applying the bound to the square-root remainder
proves Theorem~\ref{thm:main}.

\subsection{A conditional second-moment estimate}

For $\sigma>0$ and $A\ge0$, define the sandwiched quasi-relative entropy
of order two by
\begin{equation}\label{eq:weighted-second-moment}
 Q_2(A\|\sigma)=\|\sigma^{-1/4}A\sigma^{-1/4}\|_2^2,
\end{equation}
where $\|\cdot\|_2$ is the Hilbert--Schmidt norm for the ordinary trace.

\begin{lemma}\label{lem:dual-moment}
Let $E:B(\mathcal H)\to \mathcal{N}$ be a faithful conditional expectation onto $\mathcal N$, and let $\Phi$ be
unital and completely positive with $\Phi E=E\Phi=E$.
Suppose that, for some $2<q<\infty$ and $M\ge1$,
\[
 \|\Phi:L_2(\omega_0)\longrightarrow
 L_2^q(\mathcal N,\omega_0)\|\le M.
\]
Let $\sigma>0$ satisfy $E_*\sigma=\sigma$.
If $A\ge0$ and $c,\eta>0$ satisfy
\[
 A\le\eta\sigma,\qquad E_*A\le c\sigma,
\]
then
\begin{equation}\label{eq:conditional-second-moment}
 Q_2(\Phi_*A\|\sigma)
 \le M^2c^{1-2/q}\eta^{2/q}\Tr A.
\end{equation}
\end{lemma}

\begin{proof}
Put $q'=q/(q-1)$.
The duality of the conditional spaces
\cite[Proposition~5.2(ii)]{BCG} gives
\begin{equation}\label{eq:dual-hb}
 \|\Phi^\sharp:L_2^{q'}(\mathcal N,\omega_0)
                   \longrightarrow L_2(\omega_0)\|\le M,
 \qquad
 \Phi^\sharp=\Gamma_{\omega_0}^{-1}\Phi_*\Gamma_{\omega_0}.
\end{equation}
Write $\sigma=\omega_0a$ as in~\eqref{eq:invariant-density}, and set
\[
 B=a^{-1/4}Aa^{-1/4},\qquad
 E_*B=\omega_0b.
\]
Here $a>0$ and $b\ge0$ belong to $\mathcal N$.
Then
\begin{align*}
 Q_2(\Phi_*A\|\sigma)
 =\|\Phi^\sharp(\Gamma_{\omega_0}^{-1}(B))\|_{2,\omega_0}^2\le M^2\|\Gamma_{\omega_0}^{-1}(B)\|_{L_2^{q'}(\mathcal N,\omega_0)}^2.
\end{align*}
The equality uses $a\omega_0=\omega_0a$ and the bimodularity identity
$\Phi_*B=a^{-1/4}\Phi_*A a^{-1/4}$; the inequality
is~\eqref{eq:dual-hb}.
We estimate $\|\Gamma_{\omega_0}^{-1}(B)\|_{L_2^{q'}(\mathcal N,\omega_0)}^2$
by interpolation.
Take all suprema below over $R,S\in\mathcal N$ with
$\|R\|_{4,\omega_0},\|S\|_{4,\omega_0}\le1$. Then
\begin{align*}
 \|\Gamma_{\omega_0}^{-1}(B)\|_{L_2^1(\mathcal N,\omega_0)}^2
 &=\sup_{R,S}\|R\Gamma_{\omega_0}^{-1}(B)S\|_{1,\omega_0}^2\\
 &=\sup_{R,S}\bigl\|\omega_0^{1/2}R\omega_0^{-1/2}
           B\omega_0^{-1/2}S\omega_0^{1/2}\bigr\|_1^2\\
 &=\sup_{R,S}\|RBS\|_1^2\\
 &\le\sup_{R,S}(\Tr BR^*R)(\Tr BSS^*)\\
 &=\sup_{R,S}(\Tr\omega_0bR^*R)(\Tr\omega_0bSS^*)\\
 &\le\|b\|_{2,\omega_0}^2\le c\Tr A.
\end{align*}
The estimate of $\|RBS\|_1$ follows from the Cauchy--Schwarz inequality, and the trace
identity uses $E_*B=\omega_0b$.
Since $\Tr\omega_0(\,\cdot\,)=\tau_d$ on $\mathcal N$, $\omega_0$-weighted Cauchy--Schwarz and the bounds on $R,S$ give the estimate by
$\|b\|_{2,\omega_0}^2$.
For the last inequality, the marginal condition $E_*A\le c\sigma$ gives
$b\le ca^{1/2}$, while trace duality gives
$$\Tr\omega_0b^2\le  c\Tr\omega_0ba^{1/2}=c\Tr Ba^{1/2}=c\Tr A.$$

For the other endpoint,
\begin{align*}
 \|\Gamma_{\omega_0}^{-1}(B)\|_{2,\omega_0}^2=\Tr\bigl[(\sigma^{-1/4}A\sigma^{-1/4})^2\bigr]\le\eta\Tr\bigl[(\sigma^{-1/4}A\sigma^{-1/4})\sigma^{1/2}\bigr]=\eta\Tr A.
\end{align*}
The inequality follows from
$0\le\sigma^{-1/4}A\sigma^{-1/4}\le\eta\sigma^{1/2}$.
Complex interpolation between $L_2^1$ and $L_2^2$, with parameter
$2/q$ \cite[Proposition~5.2(iv)]{BCG}, now gives
\begin{align*}
 Q_2(\Phi_*A\|\sigma)
 &\le M^2\|\Gamma_{\omega_0}^{-1}(B)\|_{L_2^{q'}(\mathcal N,\omega_0)}^2\\
 &\le M^2(c\Tr A)^{1-2/q}(\eta\Tr A)^{2/q}=M^2c^{1-2/q}\eta^{2/q}\Tr A. \qedhere
\end{align*}
\end{proof}

\subsection{Conditional square-root truncation}

The next lemma constructs a square-root decomposition whose low part
satisfies both hypotheses of Lemma~\ref{lem:dual-moment}. A positive
contraction in $\mathcal N$ repairs the marginal of an initial
truncation, with an error controlled by the input tail.

\begin{lemma}\label{lem:conditional-truncation}
Let $E$ be a faithful conditional expectation, let $\sigma>0$
satisfy $E_*\sigma=\sigma$, and let $A\ge0$ satisfy
$E_*A\le4\sigma$. For every $c>4$ and $\eta>0$, there is a
decomposition $\sqrt A=L_A+H_A$ such that
\begin{equation}\label{eq:conditional-low-part}
 L_A L_A^*\le\eta\sigma,\quad
 E_*(L_A L_A^*)\le c\sigma,\quad
 \|L_A\|_2^2\le\Tr A,
\end{equation}
and
\begin{equation}\label{eq:conditional-high-part}
 \|H_A\|_2^2\le\frac c{c-4}e_\eta(A\mid\sigma).
\end{equation}
\end{lemma}

\begin{proof}
Set
\[
 R=\eta\sigma+(A-\eta\sigma)_+,\qquad B=R^{-1/2}\sqrt A.
\]
Write
\[
 \sqrt A=L+H,\qquad
 L=\sqrt{\eta\sigma}\,B,\quad H=(\sqrt R-\sqrt{\eta\sigma})B.
\]
Since $R\ge A,R\ge \eta\sigma$, we have $BB^*\le I$.
It follows that $LL^*\le\eta\sigma$ and
\[
 \|L\|_2^2=\eta\Tr A R^{-1/2}\sigma R^{-1/2}\le\Tr A.
\]
Since $\sqrt R\ge\sqrt{\eta\sigma}$,
\begin{align*}
 \|H\|_2^2
 &\le\Tr(\sqrt R-\sqrt{\eta\sigma})^2\\
 &=\Tr(A-\eta\sigma)_+
   -2\Tr\sqrt{\eta\sigma}(\sqrt R-\sqrt{\eta\sigma})\\
 &\le e_\eta(A\mid\sigma).
\end{align*}

To estimate the marginal of $LL^*$, set
\[
 S=c\sigma+(E_*(LL^*)-c\sigma)_+,
\]
so that $S\ge E_*(LL^*), S\ge c\sigma$, and define
\begin{equation}\label{eq:repair-filter}
 G=S^{-1/2}(S^{1/2}c\sigma S^{1/2})^{1/2}S^{-1/2}.
\end{equation}
Then $GSG=c\sigma$. To see that $G\in\mathcal N$, write
$\sigma=\omega_0a$ and $E_*(LL^*)=\omega_0b$ with $a,b\in\mathcal N_+$,
$a>0$. Since $\omega_0\in\mathcal N'$, we have
$S=\omega_0v$, where $v=ca+(b-ca)_+$, and
\[
 G=v^{-1/2}(v^{1/2}ca v^{1/2})^{1/2}v^{-1/2}\in\mathcal N.
\]
Note that $0<G\le I$.
From $S\ge c\sigma, S\ge E_*(LL^*)$ we obtain
$G\sigma G\le\sigma$ and $E_*(GLL^*G)=GE_*(LL^*)G\le c\sigma$.
Put $L_A=GL$ and $H_A=\sqrt A-GL$. Then
\[
 L_A L_A^*\le\eta G\sigma G\le\eta\sigma,\qquad
 \|L_A\|_2^2\le\|L\|_2^2\le\Tr A,
\]
which completes~\eqref{eq:conditional-low-part}.

For the remainder, we have
\begin{align*}
 \|H_A\|_2^2=\|\sqrt A-GL\|_2^2
 &\le\Tr\sqrt A(I-G)\sqrt A+\Tr H^*GH\\
 &\le4\Tr\sigma(I-G)+\|H\|_2^2.
\end{align*}
The first inequality follows by expanding the square and using
$G^2\le G$ and $H=\sqrt A-L$; the last follows from trace duality,
$G\in\mathcal N$, $E_*A\le4\sigma$, and $G\le I$.
On the other hand, taking traces in $GSG=c\sigma$ gives
\[
 c\Tr\sigma(I-G^2)=  \Tr (SG^2-c\sigma G^2)
 =\Tr\bigl[(E_*(LL^*)-c\sigma)_+G^2\bigr]
 \le\Tr(E_*(LL^*)-c\sigma)_+.
\]
Expanding $LL^*$ and applying the weighted operator Young inequality
to the cross terms gives
\begin{align*}
 LL^*
 &=A+HH^*-\sqrt A H^*-H\sqrt A\\
 &\le A+HH^*+\frac{c-4}{4}A+\frac4{c-4}HH^*\\
 &=\frac c4 A+\frac c{c-4}HH^*.
\end{align*}
Apply $E_*$ to obtain
\begin{equation}\label{eq:repair-excess}
 \Tr(E_*(LL^*)-c\sigma)_+\le\frac c{c-4}\|H\|_2^2.
\end{equation}
Since $I-G\le I-G^2$, the above estimates combined give
\[
 \|H_A\|_2^2
 \le\left(\frac4{c-4}+1\right)\|H\|_2^2
 \le\frac c{c-4}e_\eta(A\mid\sigma),
\]
which proves~\eqref{eq:conditional-high-part}.
\end{proof}

\subsection{Integrated tails and entropy contraction}

We combine the conditional truncation with
Lemma~\ref{lem:dual-moment} to bound the output hockey-stick
tail. The second-moment estimate controls the truncated input,
while the input hockey-stick tail controls the error. Integrating
the resulting estimate gives the contraction below.

For $q>2$, set $b=1-2/q$ and define
\begin{equation}\label{eq:explicit-tail-parameters}
 \begin{gathered}
 \vartheta_q=\frac{q+14}{4(q+2)},\qquad
 C_q=\frac{4(q+30)}{q-2},\qquad
 \gamma_q=\frac{q-2}{q+30},\\
 T_q(M)=\frac{M^2C_q^b(1+\gamma_q^{-1})^2}{2b},
 \qquad \mathcal B_q(M)=1+\log T_q(M).
 \end{gathered}
\end{equation}
Here $0<\vartheta_q<1$ and $T_q(M)>1$.

\begin{lemma}\label{lem:integrated-tail}
Let $E:B(\mathcal H)\to \mathcal N$ be a faithful conditional expectation onto $\mathcal N $ and let $\Phi$ be
unital and completely positive with $\Phi E=E\Phi=E$.
Suppose that, for some $2<q<\infty$ and $M\ge1$,
\[
 \|\Phi:L_2(\omega_0)\longrightarrow
 L_2^q(\mathcal N,\omega_0)\|\le M.
\]
Let $\vartheta_q$ and $\mathcal B_q(M)$ be defined
by~\eqref{eq:explicit-tail-parameters}.
For every $\sigma\ge0$ satisfying $E_*\sigma=\sigma$ and every
$A\ge0$ satisfying $E_*A\le4\sigma$,
\begin{equation}\label{eq:explicit-integrated-tail}
 F(\Phi_*A\mid\sigma)
 \le\vartheta_qF(A\mid\sigma)+\mathcal B_q(M)\Tr A.
\end{equation}
\end{lemma}

\begin{proof}
First suppose that $\sigma>0$, and put $T=T_q(M)$.
We split the tail integral as
\[
 F(\Phi_*A\mid\sigma)
 =\int_1^T e_t(\Phi_*A\mid\sigma)\,\frac{dt}{t}
  +\int_T^\infty e_t(\Phi_*A\mid\sigma)\,\frac{dt}{t}.
\]
Since $\Phi_*$ preserves the trace,
$e_t(\Phi_*A\mid\sigma)\le\Tr\Phi_*A=\Tr A$, so the first integral
is at most $(\Tr A)\log T$.

To estimate the second integral, recall that, for $X\ge0$,
\[
 Q_2(X\|\sigma)
 =\Tr\bigl[(\sigma^{-1/4}X\sigma^{-1/4})^2\bigr].
\]
For every $t>0$, we have
\begin{equation}\label{eq:second-moment-tail}
 e_t(X\mid\sigma)\le\frac{Q_2(X\|\sigma)}{4t}.
\end{equation}
Indeed, for $0\le K\le I$,
\begin{align*}
 \Tr K(X-t\sigma)
 &\le\sqrt{Q_2(X\|\sigma)\Tr\sigma^{1/2}K\sigma^{1/2}K}
       -t\Tr\sigma K\\
 &\le\sqrt{Q_2(X\|\sigma)\Tr\sigma K^2}
       -t\Tr\sigma K\\
 &\le\sqrt{Q_2(X\|\sigma)\Tr\sigma K}-t\Tr\sigma K
 \le\frac{Q_2(X\|\sigma)}{4t}.
\end{align*}
The first inequality is the Hilbert--Schmidt Cauchy--Schwarz inequality. The next two
use $\|\sigma^{1/2}K-K\sigma^{1/2}\|_2^2\ge0$ and $K^2\le K$,
respectively; the last follows by completing the square.
Taking the supremum over $K$ proves~\eqref{eq:second-moment-tail}.

Fix $C>4$ and $\gamma,\eta>0$. Apply
Lemma~\ref{lem:conditional-truncation} with $c=C$ to obtain
$\sqrt A=L_A+H_A$ satisfying~\eqref{eq:conditional-low-part}
and~\eqref{eq:conditional-high-part}. Young's inequality yields
\[
 A\le(1+\gamma^{-1})L_A L_A^*
 +(1+\gamma)H_A H_A^*.
\]
The variational formula~\eqref{eq:hockey-stick} gives
$e_t(X+Y\mid\sigma)\le e_t(X\mid\sigma)+\Tr Y$ for $X,Y\ge0$.
Applying these to the output of the preceding inequality gives
\begin{align}
 e_t(\Phi_*A\mid\sigma)
 &\le(1+\gamma)\Tr \Phi_*(H_A H_A^*)
 +e_t\bigl((1+\gamma^{-1})\Phi_*(L_A L_A^*)\mid\sigma\bigr)\notag\\
 &\le(1+\gamma)\|H_A\|_2^2
 +\frac{(1+\gamma^{-1})^2}{4t}
   Q_2(\Phi_*(L_A L_A^*)\|\sigma)\notag\\
 &\le(1+\gamma)\|H_A\|_2^2
 +\frac{(1+\gamma^{-1})^2M^2C^b}{4t}\eta^{2/q}
   \Tr(L_A L_A^*)\notag\\
 &\le(1+\gamma)\frac C{C-4}e_\eta(A\mid\sigma)
 +\frac{(1+\gamma^{-1})^2M^2C^b}{4t}\eta^{2/q}\Tr A.
 \label{eq:conditional-pointwise-tail}
\end{align}
The second uses~\eqref{eq:second-moment-tail} and
the quadratic homogeneity of $Q_2$.
The third inequality follows from Lemma~\ref{lem:dual-moment}
with $c=C$ and the first two bounds in~\eqref{eq:conditional-low-part}.
The last uses the trace bound in~\eqref{eq:conditional-low-part}
and~\eqref{eq:conditional-high-part}.

Now take $C=C_q$, $\gamma=\gamma_q$, and $a=(q+2)/4$.
These choices give
\[
 \frac{(1+\gamma)C}{a(C-4)}=\vartheta_q,\qquad
 1-\frac{2a}{q}=\frac b2.
\]
On $t\ge T=T_q(M)$, set $\eta=(t/T)^a$.
The first term of~\eqref{eq:conditional-pointwise-tail} integrates to
\[
 (1+\gamma)\frac C{C-4}
 \int_T^\infty e_{(t/T)^a}(A\mid\sigma)\,\frac{dt}{t}
 =\vartheta_qF(A\mid\sigma).
\]
The integral of its second term is
\[
 \frac{(1+\gamma^{-1})^2M^2C^b}{4}\Tr A
 \int_T^\infty(t/T)^{2a/q}\,\frac{dt}{t^2}
 =\frac{(1+\gamma^{-1})^2M^2C^b}{2bT}\Tr A=\Tr A.
\]
Adding the three contributions gives
\begin{align*}
 F(\Phi_*A\mid\sigma)
 &\le(\Tr A)\log T+\vartheta_qF(A\mid\sigma)+\Tr A\\
 &=\vartheta_qF(A\mid\sigma)+\mathcal B_q(M)\Tr A,
\end{align*}
which proves~\eqref{eq:explicit-integrated-tail} when $\sigma>0$.

For singular $\sigma$, replace it by
$\sigma_\epsilon=\sigma+\epsilon\omega_0$.
These positive matrices remain $E_*$-invariant and satisfy
$E_*A\le4\sigma_\epsilon$.
As $\epsilon\downarrow0$, the positive-part traces increase to
those with reference $\sigma$. Monotone convergence in the
definition of $F$ proves the assertion.
\end{proof}

\begin{proof}[Proof of Theorem~\ref{thm:main}]
By~\eqref{eq:hb-change-weight} and the faithful-state converse in
Lemma~\ref{lem:gns-transfer}, the HB and GNS contraction bounds hold
at $\omega_0$ with the same $M,s$.
For any density matrix $\rho$, Lemma~\ref{lem:sqrt-comparison} gives
$E_*W_\rho\le4E_*\rho$.
Applying Lemma~\ref{lem:integrated-tail} with $A=W_\rho$ verifies
\eqref{eq:abstract-tail-contraction} for every $\rho$, with
$\theta=\vartheta_q$ and $B=\mathcal B_q(M)$.
Proposition~\ref{prop:closure} therefore proves~\eqref{eq:main-ec} with
\begin{equation}\label{eq:explicit-ec}
 \varepsilon_q(M,s)=
 \frac{\kappa(1-\vartheta_q)}
 {\kappa+4K_0+(8/\pi)
       \bigl(\mathcal B_q(M)+\vartheta_q\log4\bigr)}.
\end{equation}
Here $\kappa=(1-s^2)\le1$ and
$\mathcal B_q(M)=\mathcal B_q(1)+2\log M$. Then
the bound~\eqref{eq:fixed-q-summary} holds with
\begin{equation}\label{eq:cq-explicit}
 c(q)=
 \frac{1-\vartheta_q}
 {1+4K_0+(8/\pi)\bigl(\mathcal B_q(1)+\vartheta_q\log4+2\bigr)}>0.
\end{equation}
As $q\downarrow2$, $1-\vartheta_q\sim3(q-2)/16$ and
$\mathcal B_q(1)\sim3\log(1/(q-2))$, so this choice satisfies
$c(q)\sim\frac{\pi}{128}\frac{q-2}{\log(1/(q-2))}$.
This finishes the proof.
\end{proof}

\begin{remark}\label{rem:fixed-time-iteration}\label{cor:qms-iteration}
Let $T_t$ be a QMS compatible with the conditional expectation $E=ET_t=T_tE$. If $T_{t_0}$ satisfies the
hypotheses of Theorem~\ref{thm:main} at some $t_0>0$, then
\begin{equation}\label{eq:iteration}
 D(T_{t*}\rho\|E_*\rho)
 \le(1-\varepsilon)^{\lfloor t/t_0\rfloor}D(\rho\|E_*\rho),
\end{equation}
for every $t\ge0$ and density matrix $\rho$.
Apply the single-step estimate at each full step and data processing
to the remaining time interval. The bound is non-strict for $0<t<t_0$,
so fixed-step iteration alone does not establish a positive MLSI constant.
\end{remark}

\section{Entropy production and modified logarithmic Sobolev inequalities}\label{sec:semigroup-general}
In this section, we prove Theorem~\ref{thm:kms-cmlsi}, which derives modified logarithmic Sobolev inequalities from conditional hyperboundedness under KMS symmetry. We first show that a small-time limit in Proposition~\ref{prop:finite-step-comparison} gives a universal comparison between entropy production and square-root energy. Under KMS symmetry, a single-time conditional hyperboundedness estimate supplies a hyperboundedness curve via interpolation.
These estimates, combined with GNS
coercivity, imply a modified
logarithmic Sobolev inequality. 

\subsection{Entropy production and square-root energy}
\label{subsec:general-energy}

Let $T_t=e^{t\mathcal L}$ be a quantum Markov semigroup on
$B(\mathcal H)$ with a faithful invariant state $\sigma$.
In this subsection, $\sigma$ need not be the canonical choice $\omega_0$. Using
$\Gamma_\sigma^{1/2}$ from~\eqref{eq:static-kms-implementation}, define their Hilbert--Schmidt implementations
\begin{equation}\label{eq:kms-implementation}
 \widehat{\mathcal L}=\Gamma_\sigma^{1/2}\mathcal L \Gamma_\sigma^{-1/2},\qquad
 \widehat T_t=\Gamma_\sigma^{1/2} T_t\Gamma_\sigma^{-1/2}.
\end{equation}
The real part of its energy is
\begin{equation}\label{eq:real-energy}
 \mathcal E_{\mathrm{re}}(A)
 =-\Re\Tr A^*\widehat{\mathcal L}(A),
 \qquad A\in B(\mathcal H).
\end{equation}
We do not assume that $\widehat{\mathcal L}$ is self-adjoint.
For a faithful state $\rho$, define its entropy production
relative to $\sigma$ by
\begin{equation}\label{eq:kms-entropy-production}
 \mathcal I_\sigma(\rho)
 =-\Tr\mathcal L_*(\rho)(\log\rho-\log\sigma).
\end{equation}
Note that both the energy and entropy production are independent of the choice of the faithful invariant state.
Under KMS symmetry, $\widehat{\mathcal L}$ is Hilbert--Schmidt
self-adjoint, and we write
\begin{equation}\label{eq:kms-energy}
 \widehat{\mathcal E}(A)
 =-\Tr A^*\widehat{\mathcal L}(A)
 =\mathcal E_{\mathrm{re}}(A).
\end{equation}

\begin{theorem}\label{thm:general-comparison}\label{thm:kms-comparison}
Let $T_t=e^{t\mathcal L}$ be a quantum Markov semigroup on
$B(\mathcal H)$ with a faithful invariant state $\sigma$.
For every faithful state $\rho$,
\begin{equation}\label{eq:general-comparison}
 \mathcal I_\sigma(\rho)
 \ge\frac\pi2\mathcal E_{\mathrm{re}}(\sqrt\rho).
\end{equation}
If the semigroup is KMS-symmetric with respect to $\sigma$, then
\begin{equation}\label{eq:kms-universal-comparison}
 \mathcal I_\sigma(\rho)
 \ge\pi\,\widehat{\mathcal E}(\sqrt\rho).
\end{equation}
\end{theorem}

\begin{proof}
For $h>0$, apply Proposition~\ref{prop:finite-step-comparison}
to $\Phi=T_h$ with $\tau=\rho$. The left-hand side is
\begin{align*}
 &D(\rho\|\sigma)-D(T_{h*}\rho\|\sigma)
   +D(T_{h*}\rho\|\rho)\\
 &\qquad=-\Tr(T_{h*}\rho-\rho)(\log\rho-\log\sigma)
 =h\mathcal I_\sigma(\rho)+o(h).
\end{align*}
Since $\widehat T_h^{\,*}
=\Gamma_\sigma^{-1/2}T_{h*}\Gamma_\sigma^{1/2}$,
Hilbert--Schmidt duality and~\eqref{eq:real-energy} give
\[
 a_{\rho,h}
 =\langle\sqrt\rho,\widehat T_h^{\,*}\sqrt\rho\rangle_{\mathrm{HS}}
 =1-h\mathcal E_{\mathrm{re}}(\sqrt\rho)+o(h),
\]
where the adjoint is for the Hilbert--Schmidt inner product. Hence
\[
 \frac{1-a_{\rho,h}}{1+a_{\rho,h}}
 =\frac h2\mathcal E_{\mathrm{re}}(\sqrt\rho)+o(h).
\]
Dividing~\eqref{eq:finite-step-comparison} by $h$ and letting
$h\downarrow0$ proves~\eqref{eq:general-comparison}.
Under KMS symmetry, $\mathcal E_{\mathrm{re}}=\widehat{\mathcal E}$,
and~\eqref{eq:finite-step-kms-comparison} doubles the coefficient,
giving~\eqref{eq:kms-universal-comparison} by the same limit.
\end{proof}

\begin{remark}\label{rem:kms-comparison-constant}
The two coefficients are uniform over finite-dimensional generators;
we do not claim optimality. In the KMS case, the counterexample in
\cite[Proposition~2.10]{BRS} rules out the coefficient $4$ in general, while the above theorem shows that the coefficient $\pi$ holds.
\end{remark}

\subsection{GNS coercivity and conditional remainders}
\label{subsec:semigroup-remainders}

For the conditional estimates below, use the canonical reference
$\omega_0=E_*(I_d/d)$ in~\eqref{eq:kms-implementation} and
the static projection $P$ from Lemma~\ref{lem:canonical-root-projection}.
The following lemma combines Wirth's GNS--KMS gap comparison
\cite{Wirth26}, the standard semigroup decay criterion, and the
corresponding energy formulation.

\begin{lemma}\label{lem:gns-kms-coercivity}
Let $E$ be a faithful conditional expectation and let
$T_t=e^{t\mathcal L}$ be a quantum Markov semigroup with
$T_tE=ET_t=E$ for  $t\ge0$.
Suppose that some $\lambda>0$ satisfies
\[
 -\Re\Tr\omega_0X^*\mathcal L(X)
 \ge\lambda\|X\|_{2,\mathrm{GNS},\omega_0}^2
 \qquad \forall \ X\in B(\mathcal H),\ EX=0.
\]
Then, for all $A\in B(\mathcal H)$ and $t\ge0$,
\begin{equation}\label{eq:semigroup-gns-contraction}
 \|(T_t-E)A\|_{2,\mathrm{GNS},\omega_0}
 \le e^{-\lambda t}\|A\|_{2,\mathrm{GNS},\omega_0}
\end{equation}
\begin{equation}\label{eq:semigroup-kms-contraction}
 \|(\widehat T_t-P)A\|_2\le e^{-\lambda t}\|A\|_2
\end{equation}
\begin{equation}\label{eq:kms-poincare}
 \mathcal E_{\mathrm{re}}(A)
 \ge\lambda\|(I-P)A\|_2^2
\end{equation}
\end{lemma}

\begin{proof}
Since $ET_h(I-E)=0$, \cite[Lemma~3.6]{Wirth26} on $\ker E$
and GNS orthogonality of $E$ give, for $A\in B(\mathcal H)$,
\begin{align*}
 \|(T_h-E)A\|_{2,\mathrm{GNS},\omega_0}
 &=\|T_h(A-EA)\|_{2,\mathrm{GNS},\omega_0}\\
 &\le e^{-\lambda h}\|A-EA\|_{2,\mathrm{GNS},\omega_0}
 \le e^{-\lambda h}\|A\|_{2,\mathrm{GNS},\omega_0}.
\end{align*}
This proves~\eqref{eq:semigroup-gns-contraction}, and hence
$T_t\to E$ as $t\to\infty$.
By \cite[Corollary~3.7 and Remark~3.9]{Wirth26} and KMS
orthogonality of $E$,
\begin{align*}
 \|(\widehat T_h-P)A\|_2
 &=\|(T_h-E)\Gamma_{\omega_0}^{-1/2}(A)\|_{2,\omega_0}\\
 &\le e^{-\lambda h}\|\Gamma_{\omega_0}^{-1/2}(A)\|_{2,\omega_0}
 =e^{-\lambda h}\|A\|_2,
\end{align*}
which proves~\eqref{eq:semigroup-kms-contraction}.
Applying~\eqref{eq:semigroup-kms-contraction} to $(I-P)A$
and using $P(I-P)=0$ gives
\[
 \|\widehat T_h(I-P)A\|_2^2
 \le e^{-2\lambda h}\|(I-P)A\|_2^2.
\]
Compatibility gives $\widehat{\mathcal L}P=P\widehat{\mathcal L}=0$.
Using the preceding estimate, we obtain~\eqref{eq:kms-poincare}:
\begin{align*}
 \mathcal E_{\mathrm{re}}(A)
 &=\mathcal E_{\mathrm{re}}((I-P)A)\\
 &=-\frac12\lim_{h\to 0}
 \frac{\|\widehat T_h(I-P)A\|_2^2-\|(I-P)A\|_2^2}{h}\\
 &\ge-\frac12\lim_{h\to 0}
 \frac{e^{-2\lambda h}-1}{h}\,\|(I-P)A\|_2^2
 =\lambda\|(I-P)A\|_2^2.\qedhere
\end{align*}
\end{proof}

For a faithful state $\rho$, recall that
$W_\rho=(\sqrt\rho-\sqrt{E_*\rho})^2$, and define the entropy production
\begin{equation}\label{eq:conditional-entropy-production}
 \mathcal I(\rho)
 =-\Tr\mathcal L_*(\rho)(\log\rho-\log E_*\rho).
\end{equation}
Under $T_tE=ET_t=E$, differentiating~\eqref{eq:kms-homogeneous-pythagoras} with
$A=T_{t*}\rho$ at $t=0$ gives the invariance under the choice of faithful invariant state
\begin{equation}\label{eq:conditional-relative-production}
 \mathcal I(\rho)=\mathcal I_{\omega_0}(\rho).
\end{equation}
The next lemma collects the conditional remainder bounds with the
coefficient $c$ of the energy comparison kept explicit. We will use
$c=\pi/2$ in general and $c=\pi$ under KMS symmetry.

\begin{lemma}\label{lem:conditional-remainder-bounds}
Let $T_t$ and $E$ satisfy the hypotheses of
Lemma~\ref{lem:gns-kms-coercivity}, with GNS coercivity constant
$\lambda>0$. Suppose that there is a constant $c>0$ such that for every faithful state $\rho$
\begin{equation}\label{eq:entropy-energy-assumption}
 \mathcal I_{\omega_0}(\rho)
 \ge c\,\mathcal E_{\mathrm{re}}(\sqrt\rho).
\end{equation} Then every such $\rho$
satisfies
\begin{equation}\label{eq:conditional-remainder-bounds}
 D\left(E_*\rho\middle\|\frac{\rho+E_*\rho}{2}\right)
 \le\frac{\mathcal I(\rho)}\lambda,
 \qquad
 \mathcal E_{\mathrm{re}}(\sqrt{W_\rho})\le\frac{\mathcal I(\rho)}c,
 \qquad
 \Tr W_\rho\le\frac{2\mathcal I(\rho)}{c\lambda}.
\end{equation}
\end{lemma}

\begin{proof}
Apply~\eqref{eq:midpoint-budget} to $T_h$, using
\eqref{eq:semigroup-gns-contraction}, and let $h\to 0$.
This gives
\[
 D\left(E_*\rho\middle\|\frac{\rho+E_*\rho}{2}\right)
 \le\lim_{h\to 0}
 \frac{2(D(\rho\|E_*\rho)-D(T_{h*}\rho\|E_*\rho))}
      {1-e^{-2\lambda h}}
 =\frac{\mathcal I(\rho)}\lambda.
\]

Set $A_\pm=(\sqrt\rho-\sqrt{E_*\rho})_\pm$, so that
$\sqrt{W_\rho}=\lvert\sqrt\rho-\sqrt{E_*\rho}\rvert=A_++A_-$.
Positivity of $\widehat T_h$ gives
\begin{align*}
 &\Tr\sqrt{W_\rho}\widehat T_h(\sqrt{W_\rho})
 -\Tr(\sqrt\rho-\sqrt{E_*\rho})
       \widehat T_h(\sqrt\rho-\sqrt{E_*\rho})\\
 &=2\Tr A_+\widehat T_h(A_-)+2\Tr A_-\widehat T_h(A_+)\ge0.
\end{align*}
At $h=0$, both cross terms vanish because $A_+A_-=0$.
Differentiating at zero proves the inequality in
\begin{equation}\label{eq:semigroup-absolute-value-energy}
 \mathcal E_{\mathrm{re}}(\sqrt{W_\rho})
 \le\mathcal E_{\mathrm{re}}(\sqrt\rho-\sqrt{E_*\rho})
 =\mathcal E_{\mathrm{re}}(\sqrt\rho).
\end{equation}
For the last equality, both $\widehat T_h$ and $\widehat T_h^*$ fix
$\sqrt{E_*\rho}$, since $\sqrt{E_*\rho}\in\operatorname{Ran}P$,
$\widehat T_hP=P\widehat T_h=P$, and $P=P^*$.
Thus $\widehat{\mathcal L}(\sqrt{E_*\rho})
=\widehat{\mathcal L}^{\,*}(\sqrt{E_*\rho})=0$.
Equations~\eqref{eq:entropy-energy-assumption}
and~\eqref{eq:conditional-relative-production}
complete the proof of the second bound. Finally,
\eqref{eq:projection-root-mass} and~\eqref{eq:kms-poincare}
give
\[
 \Tr W_\rho\le\frac2\lambda\mathcal E_{\mathrm{re}}(\sqrt\rho)
 \le\frac{2\mathcal I(\rho)}{c\lambda},
\]
as desired.
\end{proof}

\subsection{Hyperboundedness and entropy decay}
\label{subsec:hb-curves}

The next lemma applies the conditional Gross differentiation formula
of Bardet and Rouz\'e~\cite[Section~3.2]{BR} to a small-time HB curve.

\begin{lemma}\label{lem:curve-defective-lsi}
Let $E$ be a faithful conditional expectation and let
$T_t=e^{t\mathcal L}$ be a quantum Markov semigroup with
$T_tE=ET_t=E$ for all $t\ge0$.
Suppose that $\lambda>0$ satisfies
\[
 -\Re\Tr\omega_0X^*\mathcal L(X)
 \ge\lambda\|X\|_{2,\mathrm{GNS},\omega_0}^2
 \qquad , \ X\in B(\mathcal H),\ EX=0.
\]
Suppose that, for some $h_0>0$, functions $2<q(h)<\infty$ and
$M(h)\ge1$ satisfy
\begin{equation}\label{eq:small-time-hb-curve}
 \|T_h:L_2(\omega_0)\longrightarrow
 L_2^{q(h)}(\mathcal N,\omega_0)\|\le M(h)
 \qquad ,  \  0<h<h_0.
\end{equation}
Assume also that, as $h\to 0$,
\begin{equation}\label{eq:small-time-hb-slopes}
 1-\frac2{q(h)}=\beta h+o(h),\qquad
 \log M(h)=\eta h+o(h),
 \qquad \beta>0,\quad\eta\ge0.
\end{equation}
Then every $A\ge0$ satisfies
\begin{equation}\label{eq:curve-defective-lsi}
 D(A^2\|E_*A^2)
 \le\frac2\beta\mathcal E_{\mathrm{re}}(A)
       +\frac{2\eta}\beta\Tr A^2.
\end{equation}
\end{lemma}

\begin{proof}
To apply the conditional norm differentiation formula without assuming
smoothness of $q(h)$, fix $0<\beta'<\beta$ and set $p(h)=2/(1-\beta'h)$.
Since $1-2/p(h)=\beta'h$, the assumed asymptotics give
$p(h)\le q(h)$ for sufficiently small $h>0$.
Monotonicity of the conditional norms
\cite[Proposition~3.2(iii)]{BR} therefore gives
\[
 \|T_hX\|_{L_2^{p(h)}(\mathcal N,\omega_0)}
 \le\|T_hX\|_{L_2^{q(h)}(\mathcal N,\omega_0)}
 \le M(h)\|X\|_{2,\omega_0}.
\]
Let $A>0$ and $X=\Gamma_{\omega_0}^{-1/2}(A)$.
The path $T_hX$ is analytic and remains positive definite near
zero, and $p(h)$ is smooth and increasing. Also,
\eqref{eq:semigroup-gns-contraction} gives $T_t\to E$, so the
decoherence-free algebra and canonical reference in \cite{BR}
are $\mathcal N$ and $\omega_0$. The conditional norm derivative
\cite[Lemma~3.3(i), Theorem~3.4, and Corollary~3.5]{BR} gives
\[
 -\frac{\mathcal E_{\mathrm{re}}(A)}{\Tr A^2}
  +\frac{\beta'}{2\Tr A^2}D(A^2\|E_*A^2)
 =\left.\frac{d}{dh}
 \log\|T_hX\|_{L_2^{p(h)}(\mathcal N,\omega_0)}\right|_{h=0}
 \le\eta.
\]
Here the inequality follows from the HB bound and
$\log M(h)/h\to\eta$. Letting $\beta'\to \beta$ proves
\eqref{eq:curve-defective-lsi} for $A>0$. The case of general $A\ge0$ follows by approximation with $(A^2+\epsilon\omega_0)^{1/2}$ and continuity of the energy and squared norm.
\end{proof}

\begin{proposition}\label{prop:curve-mlsi}
Let $T_t$ and $E$ satisfy the hypotheses of
Lemma~\ref{lem:curve-defective-lsi}, with GNS coercivity constant
$\lambda>0$ and HB slopes $\beta>0$ and $\eta\ge0$
as in~\eqref{eq:small-time-hb-slopes}.
If there exists $c>0$ such that for every faithful state $\rho$,
\[
 \mathcal I_{\omega_0}(\rho)
 \ge c\,\mathcal E_{\mathrm{re}}(\sqrt\rho)\ .
\]
Then, for every $t\ge0$ and state $\rho$,
\begin{equation}\label{eq:curve-general-decay}
 D(T_{t*}\rho\|E_*\rho)
 \le e^{-2\alpha(c;\beta,\eta)t}D(\rho\|E_*\rho),
\end{equation}
where
\begin{equation}\label{eq:curve-general-rate}
 \alpha(c;\beta,\eta)
 =\frac{c\lambda\beta}
 {4\lambda+8\eta+10c\log3\,\beta}.
\end{equation}
\end{proposition}

\begin{proof}
Fix a faithful state $\rho$.
Lemmas~\ref{lem:conditional-remainder-bounds} and~\ref{lem:curve-defective-lsi} give
\[
 D(W_\rho\|E_*W_\rho)
 \le\frac2\beta\mathcal E_{\mathrm{re}}(\sqrt{W_\rho})
       +\frac{2\eta}\beta\Tr W_\rho
 \le\left(\frac2{c\beta}
          +\frac{4\eta}{c\beta\lambda}\right)\mathcal I(\rho).
\]
Lemma~\ref{lem:center-root-entropy} and the midpoint bound in
\eqref{eq:conditional-remainder-bounds} now yield
\begin{equation}\label{eq:curve-entropy-production-bound}
 D(\rho\|E_*\rho)
 \le\left[\frac2{c\beta}
          +\frac{4\eta}{c\beta\lambda}
          +\frac{5\log3}{\lambda}\right]\mathcal I(\rho)
 =\frac{\mathcal I(\rho)}{2\alpha(c;\beta,\eta)}.
\end{equation}
The trajectory $T_{t*}\rho$ remains faithful by positivity and invariance
of $\omega_0$, and $E_*T_{t*}\rho=E_*\rho$. 
Hence~\eqref{eq:curve-entropy-production-bound} gives
\[
 \frac{d}{dt}D(T_{t*}\rho\|E_*\rho)
 =-\mathcal I(T_{t*}\rho)
 \le-2\alpha(c;\beta,\eta)D(T_{t*}\rho\|E_*\rho).
\]
Integrating this differential inequality proves exponential decay for faithful states. The case of general $\rho$ follows by approximation with
$(1-\epsilon)\rho+\epsilon\omega_0$ as $\epsilon\to 0$.
\end{proof}

Theorem~\ref{thm:general-comparison} supplies $c=\pi/2$, so
Proposition~\ref{prop:curve-mlsi} holds without symmetry with
\begin{equation}\label{eq:curve-mlsi-rate}
 \alpha_{\mathrm{cur}}
 =\frac{\pi\lambda\beta}
 {8\lambda+16\eta+10\pi\log3\,\beta}.
\end{equation}
For every $t\ge0$ and state $\rho$, its conclusion is
\begin{equation}\label{eq:curve-mlsi-decay}
 D(T_{t*}\rho\|E_*\rho)
 \le e^{-2\alpha_{\mathrm{cur}}t}D(\rho\|E_*\rho).
\end{equation}

The conditional embedding of the identity supplies HB curves with
arbitrarily large exponent slope. Applying the same criterion gives
the following consequence.

\begin{corollary}\label{cor:zero-time-embedding}
Let $T_t$ and $E$ satisfy the hypotheses of
Lemma~\ref{lem:gns-kms-coercivity}, with GNS coercivity constant
$\lambda>0$, and set
\[
 M_\infty=\|\id:L_2(\omega_0)\longrightarrow
 L_2^\infty(\mathcal N,\omega_0)\|.
\]
Suppose that the entropy--energy comparison
\eqref{eq:entropy-energy-assumption} holds with $c>0$.
Then, for every $t\ge0$ and state $\rho$,
\begin{equation}\label{eq:zero-time-mlsi}
 \begin{gathered}
 D(T_{t*}\rho\|E_*\rho)
 \le e^{-2\alpha_0(c)t}D(\rho\|E_*\rho),\qquad
 \alpha_0(c)=\frac{\lambda}
 {(8/c)\log M_\infty+10\log3}.
 \end{gathered}
\end{equation}
One may take $c=\pi/2$ in general and $c=\pi$ under KMS symmetry.
\end{corollary}

\begin{proof}
Fix $Q>2$, put $b_Q=1-2/Q$, and let $\beta>0$.
Monotonicity of conditional norms gives
$\|\id:L_2(\omega_0)\to L_2^Q(\mathcal N,\omega_0)\|\le M_\infty$.
Interpolate this embedding with the identity on $L_2(\omega_0)$,
using \cite[Proposition~5.2(iv)]{BCG}, and apply the resulting bound
to $T_hX$. The $L_2(\omega_0)$ contractivity of $T_h$ gives
\[
 \|T_hX\|_{L_2^{q(h)}(\mathcal N,\omega_0)}
 \le M_\infty^{\beta h/b_Q}\|T_hX\|_{2,\omega_0}
 \le M_\infty^{\beta h/b_Q}\|X\|_{2,\omega_0},
 \qquad q(h)=\frac2{1-\beta h},
\]
for $0<h<b_Q/\beta$. Thus the slopes in
\eqref{eq:small-time-hb-slopes} are $\beta$ and
$\eta=\beta\log M_\infty/b_Q$. Proposition~\ref{prop:curve-mlsi}
gives entropy decay with rate constant
\[
 \alpha\left(c;\beta,\frac{\beta\log M_\infty}{b_Q}\right)
 =\frac{c\lambda}
 {4\lambda/\beta+8\log M_\infty/b_Q+10c\log3}.
\]
Apply this conclusion for each $Q,\beta$, then let $\beta\to\infty$
and $Q\to\infty$ to obtain~\eqref{eq:zero-time-mlsi}.
The interval $0<h<b_Q/\beta$ may depend on these parameters.
The two choices of $c$ follow from Theorem~\ref{thm:general-comparison}.
\end{proof}

\begin{remark}\label{rem:zero-time-index}
Define the Pimsner--Popa index by
\[
 C(E)=\inf\{C\ge1:A\le CE_*A\text{ for every }A\ge0\}.
\]
The conditional norm estimates
\cite[Propositions~5.2(iv), 5.6, and Lemma~5.7]{BCG} give
\[
 M_\infty\le\sqrt{C(E)}.
\]
Consequently, Corollary~\ref{cor:zero-time-embedding} gives the MLSI bounds
\begin{equation}\label{eq:zero-time-index-rates}
 \alpha\ge
 \begin{cases}
 \displaystyle\frac{\lambda}{(8/\pi)\log C(E)+10\log3},
     &\text{in general},\\[6pt]
 \displaystyle\frac{\lambda}{(4/\pi)\log C(E)+10\log3},
     &\text{under KMS symmetry}.
 \end{cases}
\end{equation}
The zero-time assumption here concerns the embedding of $T_0=\id$;
a bound for one positive-time map need not give the same embedding constant.

At a common GNS coercivity lower bound $\lambda$,
\cite[Corollary~1.2]{GG} gives, without symmetry and in the same
$e^{-2\alpha t}$ normalization,
\[
 \alpha\ge\frac{c_0\lambda}{1+\log C(E)},\qquad
 c_0=\frac{1-\log2}{4}.
\]
Since $8/\pi<10\log3<c_0^{-1}$, the general bound
in~\eqref{eq:zero-time-index-rates} implies
\[
 \alpha\ge\frac{\lambda}{10\log3\,(1+\log C(E))}
 >\frac{c_0\lambda}{1+\log C(E)}.
\]
Thus the same HB curve criterion recovers the logarithmic index dependence
and improves the explicit lower bound in that corollary, without
symmetry. No optimality of the constants is claimed.
\end{remark}

\subsection{Single-time hyperboundedness under KMS symmetry}\label{sec:kms}

The KMS part of Theorem~\ref{thm:general-comparison} supplies
the stronger energy comparison. We now use KMS symmetry to obtain
an HB curve from a single positive-time bound and apply the common
entropy-decay estimate.

\begin{proof}[Proof of Theorem~\ref{thm:kms-cmlsi}]
Equations~\eqref{eq:hb-change-weight} and~\eqref{eq:gns-gap-change-weight}
transfer the HB and GNS gap bounds to $\omega_0$ with the same
$M,\lambda$. By~\eqref{eq:invariant-kms-implementation}, KMS symmetry
is preserved as well.
Put $b=1-2/q$.
By~\eqref{eq:semigroup-gns-contraction}, $T_t\to E$ as $t\to\infty$,
so the decoherence-free algebra and canonical reference in \cite{BR}
are $\mathcal N$ and $\omega_0$. The KMS interpolation estimate
\cite[Proof of Theorem~4.7, Eq.~(4.5)]{BR}, applied
to~\eqref{eq:kms-single-time-hb}, gives, for $0<h<t_0$,
\begin{equation}\label{eq:kms-interpolated-curve}
 \|T_h:L_2(\omega_0)\to
          L_2^{q(h)}(\mathcal N,\omega_0)\|\le M^{h/t_0},
 \qquad q(h)=\frac2{1-bh/t_0}.
\end{equation}
Hence~\eqref{eq:small-time-hb-slopes} holds with
$\beta=b/t_0$ and $\eta=(\log M)/t_0$.
The KMS part of Theorem~\ref{thm:general-comparison} supplies $c=\pi$, since the
real-part energy is $\widehat{\mathcal E}$ under KMS symmetry.
Proposition~\ref{prop:curve-mlsi}, with $c=\pi$ and the rate
\eqref{eq:curve-general-rate}, therefore gives
\eqref{eq:kms-cmlsi-decay} with
\begin{align*}
 \widetilde{\alpha}
 &=\alpha\left(\pi;\frac b{t_0},\frac{\log M}{t_0}\right)\\
 &=\frac{b\lambda}
 {(4/\pi)\lambda t_0+(8/\pi)\log M+10\log3\,b}
 \ge\frac1{12}\frac{b\lambda}{\lambda t_0+\log M+b}=\alpha.\qedhere
\end{align*}
\end{proof}

\begin{remark}[Comparison with primitive results]\label{rem:primitive-hb-comparison}
For primitive KMS-symmetric semigroups, Temme, Pastawski, and
Kastoryano~\cite[Theorem~5 and the note after Corollary~6]{TPK14}
showed that a single-time HB estimate
$\|T_{t_0}:L_2(\sigma)\to L_q(\sigma)\|\le M$ yields HC and an $L_2$-LSI
constant $\alpha_2\ge \alpha_{\rm HB}$ in their normalization, where
\[
 \alpha_{\rm HB}=\frac{b\lambda}{2(\lambda t_0+\log M+b)},
 \qquad b=1-\frac2q,
\]
and $\lambda$ is the KMS spectral gap.
In our previous work \cite[Theorem~3.3]{GW}, we obtained the MLSI bound
$\alpha_{\rm MLSI}\ge \alpha_{\rm HB}$ directly, without $L_p$-regularity. 
Becker, Rouz\'e, and Salzmann~\cite[Theorem~1.2]{BRS} extend the
same entropy-decay bound to the primitive infinite-dimensional
KMS setting.

For comparison, use the GNS coercivity bound $\lambda$ of
Theorem~\ref{thm:kms-cmlsi} as a common KMS gap lower bound,
which is valid by Lemma~\ref{lem:gns-kms-coercivity}.
For $M>1$, the value $\widetilde{\alpha}$ obtained above satisfies
\[
 \frac{\widetilde{\alpha}}{\alpha_{\rm HB}}
 =\frac{2(\lambda t_0+\log M+b)}
 {(4/\pi)\lambda t_0+(8/\pi)\log M+10\log3\,b},
 \qquad
 \frac1{5\log3}\le\frac{\widetilde{\alpha}}{\alpha_{\rm HB}}\le\frac\pi2.
\]
The bound given by $\widetilde{\alpha}$ is equivalent, up to absolute constants,
to these primitive bounds, with no uniform numerical ordering. 
\end{remark}

\medskip
\noindent\textbf{Acknowledgments.}
Li Gao and Lijun Wang are supported in part by the National Natural
Science Foundation of China (grant no. 12401163) and the Department of
Science and Technology of Hubei Province (project nos. 2025EHA041 and
2025AFA044). The authors acknowledge the use of ChatGPT for repeated
discussions of proof ideas and details. All arguments were independently verified
by the authors.

\bibliographystyle{amsalpha}
\bibliography{complete_entropy_contraction}

@article{TPK14,
  author       = {Temme, K. and Pastawski, F. and Kastoryano, M. J.},
  title        = {Hypercontractivity of quasi-free quantum semigroups},
  journal      = {J. Phys. A: Math. Theor.},
  volume       = {47},
  number       = {40},
  year         = {2014},
  pages        = {405303},
  doi          = {10.1088/1751-8113/47/40/405303},
  eprint       = {1403.5224v2},
  archivePrefix = {arXiv},
  url          = {https://arxiv.org/abs/1403.5224v2},
  note         = {\href{https://doi.org/10.1088/1751-8113/47/40/405303}{doi:10.1088/1751-8113/47/40/405303}. Preprint: \url{https://arxiv.org/abs/1403.5224v2}}
}

@article{BCG,
  author       = {Bardet, I. and Capel, {\'A}. and Gao, L. and Lucia, A. and P{\'e}rez-Garc{\'i}a, D. and Rouz{\'e}, C.},
  title        = {Entropy decay for {Davies} semigroups of a one dimensional quantum lattice},
  journal      = {Comm. Math. Phys.},
  volume       = {405},
  year         = {2024},
  pages        = {42},
  doi          = {10.1007/s00220-023-04869-5},
  eprint       = {2112.00601v1},
  archivePrefix = {arXiv},
  url          = {https://arxiv.org/abs/2112.00601v1},
  note         = {\href{https://doi.org/10.1007/s00220-023-04869-5}{doi:10.1007/s00220-023-04869-5}. Preprint: \url{https://arxiv.org/abs/2112.00601v1}. The proposition numbering used here refers to this preprint version}
}

@misc{Bardet17,
  author       = {Bardet, I.},
  title        = {Estimating the decoherence time using non-commutative functional inequalities},
  howpublished = {\href{https://arxiv.org/abs/1710.01039v1}{arXiv:1710.01039v1}},
  year         = {2017},
  eprint       = {1710.01039v1},
  archivePrefix = {arXiv},
  url          = {https://arxiv.org/abs/1710.01039v1}
}

@article{BR,
  author       = {Bardet, I. and Rouz{\'e}, C.},
  title        = {Hypercontractivity and logarithmic {Sobolev} inequality for non-primitive quantum {Markov} semigroups and estimation of decoherence rates},
  journal      = {Ann. Henri Poincar{\'e}},
  volume       = {23},
  year         = {2022},
  pages        = {3839--3903},
  doi          = {10.1007/s00023-022-01196-8},
  eprint       = {1803.05379v3},
  archivePrefix = {arXiv},
  url          = {https://arxiv.org/abs/1803.05379v3},
  note         = {\href{https://doi.org/10.1007/s00023-022-01196-8}{doi:10.1007/s00023-022-01196-8}. Preprint: \url{https://arxiv.org/abs/1803.05379v3}}
}

@misc{BRS,
  author       = {Becker, S. and Rouz{\'e}, C. and Salzmann, R.},
  title        = {Hypercontractivity and entropy decay for infinite-dimensional quantum {Gibbs} samplers},
  howpublished = {\href{https://arxiv.org/abs/2610.06764v1}{arXiv:2610.06764v1}},
  year         = {2026},
  eprint       = {2610.06764v1},
  archivePrefix = {arXiv},
  url          = {https://arxiv.org/abs/2610.06764v1}
}

@article{Frenkel,
  author       = {Frenkel, P. E.},
  title        = {Integral formula for quantum relative entropy implies data processing inequality},
  journal      = {Quantum},
  volume       = {7},
  year         = {2023},
  pages        = {1102},
  doi          = {10.22331/q-2023-09-07-1102},
  eprint       = {2208.12194v4},
  archivePrefix = {arXiv},
  url          = {https://arxiv.org/abs/2208.12194v4},
  note         = {\href{https://doi.org/10.22331/q-2023-09-07-1102}{doi:10.22331/q-2023-09-07-1102}. Preprint: \url{https://arxiv.org/abs/2208.12194v4}}
}

@misc{GG,
  author       = {Gao, L. and Guo, J.},
  title        = {Tight entropy contraction beyond detailed balance},
  howpublished = {\href{https://arxiv.org/abs/2609.24953v1}{arXiv:2609.24953v1}},
  year         = {2026},
  eprint       = {2609.24953v1},
  archivePrefix = {arXiv},
  url          = {https://arxiv.org/abs/2609.24953v1}
}

@misc{GGBKM,
  author       = {Gao, L. and Guo, J.},
  title        = {Relative entropy decay via {BKM} coercivity for quantum {Markov} semigroups},
  howpublished = {\href{https://arxiv.org/abs/2609.15902v1}{arXiv:2609.15902v1}},
  year         = {2026},
  eprint       = {2609.15902v1},
  archivePrefix = {arXiv},
  url          = {https://arxiv.org/abs/2609.15902v1}
}

@article{GJL,
  author       = {Gao, L. and Junge, M. and LaRacuente, N.},
  title        = {{Fisher} information and logarithmic {Sobolev} inequality for matrix-valued functions},
  journal      = {Ann. Henri Poincar{\'e}},
  volume       = {21},
  year         = {2020},
  pages        = {3409--3478},
  doi          = {10.1007/s00023-020-00947-9},
  eprint       = {1807.08838v1},
  archivePrefix = {arXiv},
  url          = {https://arxiv.org/abs/1807.08838v1},
  note         = {\href{https://doi.org/10.1007/s00023-020-00947-9}{doi:10.1007/s00023-020-00947-9}. Preprint: \url{https://arxiv.org/abs/1807.08838v1}. The lemma numbering used here refers to this preprint version}
}

@article{GJLL,
  author       = {Gao, L. and Junge, M. and LaRacuente, N. and Li, H.},
  title        = {Complete positivity order and relative entropy decay},
  journal      = {Forum Math. Sigma},
  volume       = {13},
  year         = {2025},
  pages        = {e31},
  doi          = {10.1017/fms.2024.117},
  eprint       = {2209.11684v2},
  archivePrefix = {arXiv},
  url          = {https://arxiv.org/abs/2209.11684v2},
  note         = {\href{https://doi.org/10.1017/fms.2024.117}{doi:10.1017/fms.2024.117}. Preprint: \url{https://arxiv.org/abs/2209.11684v2}. The theorem numbering used here refers to this preprint version}
}

@article{GR,
  author       = {Gao, L. and Rouz{\'e}, C.},
  title        = {Complete entropic inequalities for quantum {Markov} chains},
  journal      = {Arch. Ration. Mech. Anal.},
  volume       = {245},
  number       = {1},
  year         = {2022},
  pages        = {183--238},
  doi          = {10.1007/s00205-022-01785-1},
  eprint       = {2102.04146v3},
  archivePrefix = {arXiv},
  url          = {https://arxiv.org/abs/2102.04146v3},
  note         = {\href{https://doi.org/10.1007/s00205-022-01785-1}{doi:10.1007/s00205-022-01785-1}. Preprint: \url{https://arxiv.org/abs/2102.04146v3}}
}

@misc{GW,
  author       = {Gao, L. and Wang, L.},
  title        = {Quantum entropy contraction and factorization from hypercontractivity},
  howpublished = {\href{https://arxiv.org/abs/2609.20753v2}{arXiv:2609.20753v2}},
  year         = {2026},
  eprint       = {2609.20753v2},
  archivePrefix = {arXiv},
  url          = {https://arxiv.org/abs/2609.20753v2}
}

@article{HT,
  author       = {Hirche, C. and Tomamichel, M.},
  title        = {Quantum {R{\'e}nyi} and $f$-divergences from integral representations},
  journal      = {Comm. Math. Phys.},
  volume       = {405},
  year         = {2024},
  pages        = {208},
  doi          = {10.1007/s00220-024-05087-3},
  eprint       = {2306.12343v3},
  archivePrefix = {arXiv},
  url          = {https://arxiv.org/abs/2306.12343v3},
  note         = {\href{https://doi.org/10.1007/s00220-024-05087-3}{doi:10.1007/s00220-024-05087-3}. Preprint: \url{https://arxiv.org/abs/2306.12343v3}}
}

@misc{CPS,
  author       = {Capel, {\'A}. and P{\'e}rez-Garc{\'i}a, D. and Scandi, M.},
  title        = {Modified logarithmic {Sobolev} inequality for {1D} non-commuting {Hamiltonians}},
  howpublished = {\href{https://arxiv.org/abs/2610.03683v1}{arXiv:2610.03683v1}},
  year         = {2026},
  eprint       = {2610.03683v1},
  archivePrefix = {arXiv},
  url          = {https://arxiv.org/abs/2610.03683v1}
}

@misc{CD,
  author       = {Chen, H. and Ding, Z.},
  title        = {Spectral analysis beyond the spectral gap: Rapid mixing of quantum {Gibbs} samplers in one dimension},
  howpublished = {\href{https://arxiv.org/abs/2610.05848v1}{arXiv:2610.05848v1}},
  year         = {2026},
  eprint       = {2610.05848v1},
  archivePrefix = {arXiv},
  url          = {https://arxiv.org/abs/2610.05848v1}
}

@misc{OW,
  author       = {O'Donnell, R. and Wright, J.},
  title        = {Efficient quantum tomography {II}},
  howpublished = {\href{https://arxiv.org/abs/1612.00034v1}{arXiv:1612.00034v1}},
  year         = {2016},
  eprint       = {1612.00034v1},
  archivePrefix = {arXiv},
  url          = {https://arxiv.org/abs/1612.00034v1}
}

@misc{Wirth26,
  author       = {Wirth, M.},
  title        = {The {KMS} and {GNS} spectral gap of quantum {Markov} semigroups},
  howpublished = {\href{https://arxiv.org/abs/2604.21630v1}{arXiv:2604.21630v1}},
  year         = {2026},
  eprint       = {2604.21630v1},
  archivePrefix = {arXiv},
  primaryClass = {math-ph},
  url          = {https://arxiv.org/abs/2604.21630v1}
}

@misc{CKG23,
  author       = {Chen, C.-F. and Kastoryano, M. J. and Gily{\'e}n, A.},
  title        = {An efficient and exact noncommutative quantum {Gibbs} sampler},
  howpublished = {\href{https://arxiv.org/abs/2311.09207v2}{arXiv:2311.09207v2}},
  year         = {2023},
  eprint       = {2311.09207v2},
  archivePrefix = {arXiv},
  url          = {https://arxiv.org/abs/2311.09207v2},
  note         = {Revised 14 October 2025}
}

@article{DLL25,
  author       = {Ding, Z. and Li, B. and Lin, L.},
  title        = {Efficient quantum {Gibbs} samplers with {Kubo--Martin--Schwinger} detailed balance condition},
  journal      = {Comm. Math. Phys.},
  volume       = {406},
  year         = {2025},
  pages        = {67},
  doi          = {10.1007/s00220-025-05235-3},
  eprint       = {2404.05998v5},
  archivePrefix = {arXiv},
  url          = {https://arxiv.org/abs/2404.05998v5},
  note         = {\href{https://doi.org/10.1007/s00220-025-05235-3}{doi:10.1007/s00220-025-05235-3}. Preprint: \url{https://arxiv.org/abs/2404.05998v5}}
}

@article{RSFA26,
  author       = {Rouz{\'e}, C. and {Stilck Fran{\c c}a}, D. and Alhambra, {\'A}. M.},
  title        = {Optimal quantum algorithm for {Gibbs} state preparation},
  journal      = {Phys. Rev. Lett.},
  volume       = {136},
  year         = {2026},
  pages        = {060601},
  doi          = {10.1103/lhht-svmn},
  eprint       = {2411.04885v2},
  archivePrefix = {arXiv},
  url          = {https://arxiv.org/abs/2411.04885v2},
  note         = {\href{https://doi.org/10.1103/lhht-svmn}{doi:10.1103/lhht-svmn}. Preprint: \url{https://arxiv.org/abs/2411.04885v2}}
}

@misc{BC25,
  author       = {Bergamaschi, T. and Chen, C.-F.},
  title        = {Fast mixing of quantum spin chains at all temperatures},
  howpublished = {\href{https://arxiv.org/abs/2510.08533v2}{arXiv:2510.08533v2}},
  year         = {2025},
  eprint       = {2510.08533v2},
  archivePrefix = {arXiv},
  url          = {https://arxiv.org/abs/2510.08533v2},
  note         = {Revised 23 January 2026}
}

@misc{RSF26,
  author       = {Rouz{\'e}, C. and {Stilck Fran{\c c}a}, D.},
  title        = {Rapid mixing of {Gibbs} samplers via quantum {Dobrushin--Shlosman} conditions},
  howpublished = {\href{https://arxiv.org/abs/2610.03578v1}{arXiv:2610.03578v1}},
  year         = {2026},
  eprint       = {2610.03578v1},
  archivePrefix = {arXiv},
  url          = {https://arxiv.org/abs/2610.03578v1}
}

@misc{JP,
  author       = {Junge, M. and Parcet, J.},
  title        = {Theory of amalgamated {$L_p$} spaces in noncommutative probability},
  howpublished = {\href{https://arxiv.org/abs/math/0511406v2}{arXiv:math/0511406v2}},
  year         = {2007},
  eprint       = {math/0511406v2},
  archivePrefix = {arXiv},
  url          = {https://arxiv.org/abs/math/0511406v2},
  note         = {Version 2, revised 30 July 2007. The section and chapter references used here refer to this preprint version}
}

@article{Umegaki62,
  author       = {Umegaki, H.},
  title        = {Conditional expectation in an operator algebra, {IV} (Entropy and information)},
  journal      = {Kodai Math. Sem. Rep.},
  volume       = {14},
  number       = {2},
  year         = {1962},
  pages        = {59--85},
  doi          = {10.2996/kmj/1138844604},
  url          = {https://doi.org/10.2996/kmj/1138844604},
  note         = {\href{https://doi.org/10.2996/kmj/1138844604}{doi:10.2996/kmj/1138844604}}
}

@article{HiaiPetz91,
  author       = {Hiai, F. and Petz, D.},
  title        = {The proper formula for relative entropy and its asymptotics in quantum probability},
  journal      = {Comm. Math. Phys.},
  volume       = {143},
  year         = {1991},
  pages        = {99--114},
  doi          = {10.1007/BF02100287},
  url          = {https://doi.org/10.1007/BF02100287},
  note         = {\href{https://doi.org/10.1007/BF02100287}{doi:10.1007/BF02100287}}
}

@article{Lindblad75,
  author       = {Lindblad, G.},
  title        = {Completely positive maps and entropy inequalities},
  journal      = {Comm. Math. Phys.},
  volume       = {40},
  year         = {1975},
  pages        = {147--151},
  doi          = {10.1007/BF01609396},
  url          = {https://doi.org/10.1007/BF01609396},
  note         = {\href{https://doi.org/10.1007/BF01609396}{doi:10.1007/BF01609396}}
}

@article{HRF22,
  author       = {Hirche, C. and Rouz{\'e}, C. and {Stilck Fran{\c c}a}, D.},
  title        = {On contraction coefficients, partial orders and approximation of capacities for quantum channels},
  journal      = {Quantum},
  volume       = {6},
  year         = {2022},
  pages        = {862},
  doi          = {10.22331/q-2022-11-28-862},
  eprint       = {2011.05949v4},
  archivePrefix = {arXiv},
  url          = {https://arxiv.org/abs/2011.05949v4},
  note         = {\href{https://doi.org/10.22331/q-2022-11-28-862}{doi:10.22331/q-2022-11-28-862}. Preprint: \url{https://arxiv.org/abs/2011.05949v4}}
}

@article{KT13,
  author       = {Kastoryano, M. J. and Temme, K.},
  title        = {Quantum logarithmic {Sobolev} inequalities and rapid mixing},
  journal      = {J. Math. Phys.},
  volume       = {54},
  number       = {5},
  year         = {2013},
  pages        = {052202},
  doi          = {10.1063/1.4804995},
  eprint       = {1207.3261v2},
  archivePrefix = {arXiv},
  url          = {https://arxiv.org/abs/1207.3261v2},
  note         = {\href{https://doi.org/10.1063/1.4804995}{doi:10.1063/1.4804995}. Preprint: \url{https://arxiv.org/abs/1207.3261v2}}
}

@misc{GGW26,
  author       = {Gao, L. and Guo, J. and Wang, S.},
  title        = {Complete modified logarithmic {Sobolev} inequalities for one-dimensional quantum spin chain},
  howpublished = {\href{https://arxiv.org/abs/2610.28600v1}{arXiv:2610.28600v1}},
  year         = {2026},
  eprint       = {2610.28600v1},
  archivePrefix = {arXiv},
  primaryClass = {quant-ph},
  url          = {https://arxiv.org/abs/2610.28600v1}
}

@misc{CDL26,
  author       = {Chen, H. and Ding, Z. and Lu, J.},
  title        = {Size-independent logarithmic {Sobolev} inequalities for quantum {Gibbs} samplers},
  howpublished = {\href{https://arxiv.org/abs/2610.31571v1}{arXiv:2610.31571v1}},
  year         = {2026},
  eprint       = {2610.31571v1},
  archivePrefix = {arXiv},
  primaryClass = {quant-ph},
  url          = {https://arxiv.org/abs/2610.31571v1}
}

@misc{LWWEntropic,
  author       = {Liu, Z. and Wan, J. and Wu, J.},
  title        = {Some entropic inequalities for primitive {KMS} symmetric quantum {Markov} semigroups},
  howpublished = {\href{https://arxiv.org/abs/2609.13726v1}{arXiv:2609.13726v1}},
  year         = {2026},
  eprint       = {2609.13726v1},
  archivePrefix = {arXiv},
  url          = {https://arxiv.org/abs/2609.13726v1}
}

@misc{WWCheeger,
  author       = {Wan, J. and Wu, J.},
  title        = {Quantum {Cheeger} inequalities for {KMS}-symmetric quantum {Markov} semigroups},
  howpublished = {\href{https://arxiv.org/abs/2609.21368v1}{arXiv:2609.21368v1}},
  year         = {2026},
  eprint       = {2609.21368v1},
  archivePrefix = {arXiv},
  url          = {https://arxiv.org/abs/2609.21368v1}
}

@misc{WuReverse,
  author       = {Wu, J.},
  title        = {Quantum reverse hypercontractivity for {KMS}-symmetric quantum {Markov} semigroups},
  howpublished = {\href{https://arxiv.org/abs/2610.00242v1}{arXiv:2610.00242v1}},
  year         = {2026},
  eprint       = {2610.00242v1},
  archivePrefix = {arXiv},
  primaryClass = {math-ph},
  url          = {https://arxiv.org/abs/2610.00242v1}
}

\bigskip
\begingroup
\small
\setlength{\parindent}{2em}
\setlength{\parskip}{0pt}

\indent\textsc{Li Gao}\par
\indent\textit{School of Mathematics and Statistics,
Wuhan University, Wuhan 430072, China}\par
\indent\textit{Wuhan Institute of Quantum Technology,
Wuhan 430075, China}\par
\indent\textit{E-mail address: }
\texttt{gao.li@whu.edu.cn}\par

\medskip

\indent\textsc{Lijun Wang}\par
\indent\textit{School of Mathematics and Statistics,
Wuhan University, Wuhan 430072, China}\par
\indent\textit{E-mail address: }
\texttt{lijun\_wang@whu.edu.cn}\par

\endgroup

\end{document}